\documentclass[journal]{IEEEtran}
\usepackage{hyperref}       
\usepackage{url}            
\usepackage{amsfonts}       
\usepackage{nicefrac}       
\usepackage{microtype}      
\usepackage{bbm}

\usepackage{xspace,amsmath,amssymb,amsthm,epsfig,syntonly,dsfont,pifont}
\usepackage{cite,bm,color,textcomp}
\usepackage[linesnumbered, ruled, vlined]{algorithm2e}
\usepackage{epstopdf}
\usepackage{empheq}
\usepackage{graphicx,subfig,balance}
\usepackage{float}
\usepackage{xfrac}
\DeclareMathOperator*{\argmax}{argmax}
\DeclareMathOperator*{\argmin}{argmin}

\SetKwInput{kwInit}{Init}

\newcommand*{\hermconj}{^{\mathsf{H}}}

\newcommand{\mM}{\mathcal{M}}
\newcommand{\mS}{\mathcal{S}}
\newcommand{\mG}{\mathcal{G}}

\newcommand{\mE}{\mathcal{E}}
\newcommand{\mH}{\mathcal{H}}

\newcommand{\mO}{\mathcal{O}}

\newcommand{\R}{\mathbb{R}}
\newcommand{\C}{\mathbb{C}}
\newcommand{\Prob}{\mathbb{P}}

\newcommand{\E}{\mathbb{E}}
\newcommand{\indfunc}{\mathbb{I}}

\newtheorem{remark}{\bf Remark}
\newtheorem{definition}{\bf Definition}
\newtheorem{theorem}{\bf Theorem}

\newcommand{\tabincell}[2]{\begin{tabular}{@{}#1@{}}#2\end{tabular}}

\begin{document}

\title{Multi-Frequency Joint Community Detection and Phase Synchronization}

\author{Lingda Wang, and Zhizhen Zhao\iffalse~\IEEEmembership{Member, IEEE}\fi \thanks{L. Wang and Z. Zhao are with the Coordinated Science Lab and the Department of Electrical and Computer Engineering, University of Illinois at Urbana-Champaign. Emails: {\tt\footnotesize \{lingdaw2, zhizhenz\}@illinois.edu}. Funding: This work was supported by NSF grant DMS-1854791 and Alfred P. Sloan foundation.}}



\maketitle
\begin{abstract}
This paper studies the joint community detection and phase synchronization problem on the \textit{stochastic block model with relative phase}, where each node is associated with an unknown phase angle. This problem, with a variety of real-world applications, aims to recover the cluster structure and associated phase angles simultaneously. We show this problem exhibits a \textit{``multi-frequency''} structure by closely examining its maximum likelihood estimation (MLE) formulation, whereas existing methods are not originated from this perspective. To this end, two simple yet efficient algorithms that leverage the MLE formulation and benefit from the information across multiple frequencies are proposed. The former is a spectral method based on the novel multi-frequency column-pivoted QR factorization. The factorization applied to the top eigenvectors of the observation matrix provides key information about the cluster structure and associated phase angles. The second approach is an iterative multi-frequency generalized power method, where each iteration updates the estimation in a matrix-multiplication-then-projection manner. Numerical experiments show that our proposed algorithms significantly improve the ability of exactly recovering the cluster structure and the accuracy of the estimated phase angles, compared to state-of-the-art algorithms.
\end{abstract}

\section{Introduction}
\label{sec:Intro}

\textit{Community detection} on \textit{stochastic block model} (SBM)~\cite{abbe2017community}, and \textit{phase synchronization}~\cite{singer2011angular}, are both of fundamental importance among multiple fields, such as machine learning~\cite{chen2018supervised,berahmand2022novel}, social science~\cite{girvan2002community,berahmand2018community}, and signal processing~\cite{singer2011viewing,zhao2014rotationally,zamiri2021mvdf}, to just name a few. 

\textbf{Community detection on SBM}. Consider the symmetric SBM with $N$ nodes that fall into $M$ underlying clusters of equal size $s = \sfrac{N}{M}$. SBM generates a random graph $\mG$ such that each pair of nodes $(i,j)$ are connected independently with probability $p$ if $(i,j)$ belong to the same cluster, and with probability $q$ otherwise. The goal is to recover underlying cluster structure of nodes, given the adjacency matrix $\bm{A}_{\text{SBM}}\in\{0,1\}^{N\times N}$ of the observed graph $\mG$. During the past decade, significant progress has been made on the information-theoretic threshold of the exact recovery on SBM~\cite{abbe2015exact,abbe2015community,abbe2017community}, in the regime where $p=\sfrac{\alpha\log{N}}{N}$, $q=\sfrac{\beta\log{N}}{N}$, and $\sqrt{\alpha} - \sqrt{\beta} > \sqrt{M}$. The maximum likelihood estimation (MLE) formulation of community detection on SBM
\begin{equation}
\label{eq:SBM_MLE}
\max_{\bm{H}\in\mH}\quad \left\langle\bm{A}_{\text{SBM}},  \bm{H}\bm{H}^\top\right\rangle, 
\end{equation}
is capable of achieving the exact recovery in the above regime, where $\mH:=\{\bm{H}\in\{0,1\}^{N\times M}: \bm{H}\bm{1}_M=\bm{1}_N,\bm{H}^\top\bm{1}_N = s\bm{1}_M\}$ is the feasible set. However, the MLE~\eqref{eq:SBM_MLE} is non-convex and NP-hard in the worst case. Therefore, different approaches based on MLE~\eqref{eq:SBM_MLE} or other formulations are proposed to tackle this problem, such as spectral method~\cite{abbe2020entrywise,krzakala2013spectral,massoulie2014community,ng2001spectral,vu2018simple,yun2014accurate,su2019strong,mcsherry2001spectral}, semidefinite programming (SDP)~\cite{abbe2015exact,amini2018semidefinite,bandeira2018random,guedon2016community,hajek2016achieving,hajek2016achievingb,perry2017semidefinite,fei2018exponential,li2021convex}, and belief propagation~\cite{abbe2015community,decelle2011asymptotic}.

\textbf{Phase synchronization}. The phase synchronization problem concerns recovering phase angles $\theta_1,\ldots,\theta_N$ in $[0,2\pi)$ from a subset of possibly noisy phase transitions $\theta_{ij} := (\theta_i-\theta_j) \mod 2\pi$. The phase synchronization problem can be encoded into an observation graph $\mG$, where each phase angle is associated with a node $i$ and the phase transitions are observed between $\theta_i$ and $\theta_j$ if and only if there is an edge in $\mG$ connecting the pair of nodes ($i, j$). Under the random corruption model~\cite{singer2011angular, chen2014information}, observations constitute a Hermitian matrix whose $(i,j)$th entry for any $i<j$ satisfies,
\begin{align*}
    \bm{A}_{\text{Ph},ij} = 
    \begin{cases}
    e^{\iota(\theta_i-\theta_j)}, &\text{with probability } r\in[0,1),\\
    u_{ij}\sim\text{Unif}(U(1)), &\text{with probability } 1-r,
    \end{cases}
\end{align*}
where $\iota = \sqrt{-1}$ is the  imaginary unit, and $U(1)$ is unitary group of dimension $1$. The most common formulation of the phase synchronization problem is through the following nonconvex optimization program
\begin{equation}
\label{eq:ps_form}
    \max_{\bm{x}\in\C_1^N}\quad \left\langle\bm{A}_{\text{Ph}}, \bm{x}\bm{x}\hermconj\right\rangle,
\end{equation}
where $\C_1^N$ is the Cartesian product of $N$ copies of $U(1)$. Again, similar to SBM, solving~\eqref{eq:ps_form} is non-convex and NP-hard~\cite{zhang2006complex}. Many algorithms have been proposed for practical and approximate solutions of~\eqref{eq:ps_form}, including spectral and SDP relaxations~\cite{singer2011angular,cucuringu2012eigenvector,chaudhury2015global,bandeira2016approximating,bandeira2017tightness}, and generalized power method (GPM)~\cite{boumal2016nonconvex,liu2017estimation, zhong2018near}. Besides, \cite{bandeira2020non,perry2018message,gao2019multi} consider the phase synchronization problem in multiple frequency channels, which in general outperforms the formulation~\eqref{eq:ps_form}.

\begin{figure}[htbp]

    \begin{center}
        \includegraphics[width = 0.9\columnwidth]{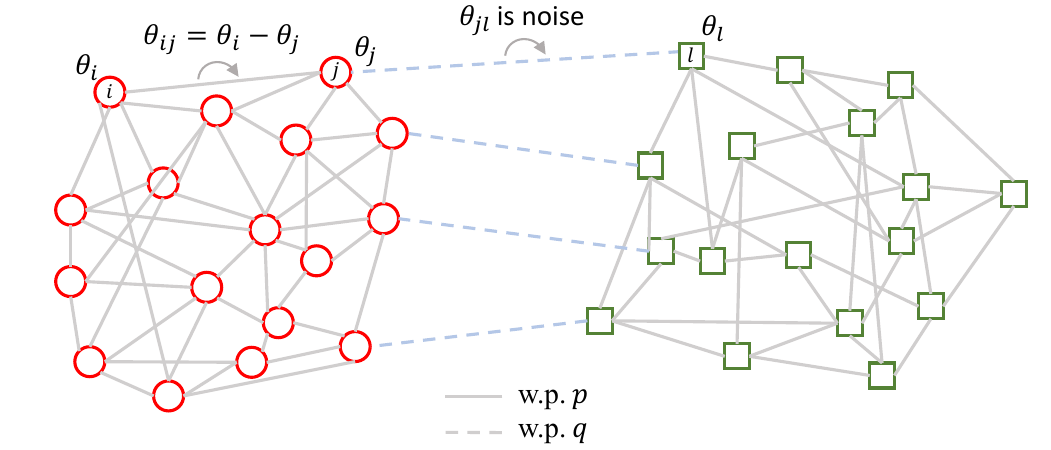}
    \end{center}
    \caption{Illustration of the joint estimation problem on a network with two clusters of equal size. Each node is associated with a phase angle. Each pair of nodes within the same cluster (resp. across clusters) are independently connected with probability $p$ (resp. $q$) as shown in solid (resp. dash) lines. Also, a phase transition $\theta_{ij}=\theta_i-\theta_j$ (resp. $\theta_{ij}$ is noise) is observed on each edge $(i,j)$ within each cluster (resp. across clusters). The goal is to recover the cluster structure and the associated phase angles simultaneously.}
    \label{fig:SBM_ph}
\end{figure}

Recently, an increasing interest~\cite{fan2022joint,fan2021spectral,chen2021non} has been seen in the \textit{joint community detection and phase (or group) synchronization problem} (joint estimation problem, for brevity). As illustrated in Figure~\ref{fig:SBM_ph}, the joint estimation problem assumes data points associated with phase angles (or group elements) in a network fall into $M$ underlying clusters, and aims to simultaneously recover the cluster structure and associated phase angles (or group elements). The joint estimation problem is motivated by the 2D class averaging procedure in \
{cryo-electron microscopy single particle reconstruction}~\cite{frank2006three,singer2011viewing,zhao2014rotationally}, which aims to cluster 2D projection images taken from similar viewing directions, align ($U(1)$ or $SO(2)$ synchronization due to the in-plane rotation) and average projection images in each cluster to improve their signal-to-noise ratio. 

In this paper, we study the joint estimation problem based on the probabilistic model, \textit{stochastic block model with relative phase} (SBM-Ph), which is similar to the probabilistic model considered in~\cite{fan2022joint,fan2021spectral,chen2021non}. Specifically, given $N$ nodes in a network assigned into $M$ underlying clusters of equal size $s=\sfrac{N}{M}$, we assume that each node $i$ is associated with an unknown phase angle $\theta^*_i\in \Omega$, where $\Omega$ is a discretization of $[0, 2\pi)$\footnote{The joint estimation problem is also extended into $[0,2\pi)$ in Section~\ref{subsec:ext}.}. For each pair of nodes $(i,j)$, if they belong to the same cluster, their phase transition $\theta_{ij} := (\theta_i - \theta_j) \mod 2\pi$ can be obtained with probability $p$; otherwise, we obtain noise generated uniformly at random from $\Omega$ with probability $q$. The goal of the joint estimation problem is to simultaneously recover the cluster structure and associated phase angles. This problem can be formulated as an optimization program maximizing not only the edge connections inside each cluster, but also the consistency among the observed phase transitions within each cluster. Still, such kind of optimization programs, similar to community detection on SBM~\eqref{eq:SBM_MLE} and phase synchronization~\eqref{eq:ps_form}, is non-convex. In~\cite{fan2022joint}, an SDP based method is proposed to achieve approximate solutions with a polynomial computational complexity. \cite{fan2021spectral} proposes a spectral method based on the block-wise column-pivoted QR (CPQR) factorization, which scales linearly with the number of edges in the network. The most recent work~\cite{chen2021non} develops an iterative GPM, where each iteration follows a matrix-multiplication-then-projection manner. The iterative GPM requires an initialization, and the computational complexity of each iteration also scales linearly with the number of edges in the network\footnote{The bottleneck of each iteration is the matrix multiplication, which is $\mO(\#\text{ of edges})$ in general. To achieve  $\mO(N\log^2{N})$ complexity claimed in~\cite{chen2021non}, one need to assume the graph is sparse.}. However, existing methods are not developed from the MLE perspective, which limit their performance on the joint estimation problem.

\subsection{Contributions}
Unlike existing methods, this paper studies the joint estimation problem by first closely examining its MLE formulation, which exhibits a \textit{``multi-frequency''} structure (detailed in Section~\ref{sec:Problem_Formulation}). More specifically, the MLE formulation is maximizing the summation over multiple frequency components, whose first frequency component is actually the objective function studied in~\cite{fan2022joint,fan2021spectral,chen2021non}. Based on the new insight, \textit{a spectral method based on the multi-frequency column-pivoted QR \textup{(MF-CPQR)} factorization} and \textit{an iterative multi-frequency generalized power method} (MF-GPM) are proposed to tackle the MLE formulation of the joint estimation problem, and both significantly outperform state-of-the-art methods in numerical experiments. The contributions of this paper can be summarized as follows:
\begin{itemize}
    \item We study the MLE formulation of the joint community detection and phase synchronization problem with discretized phase angles, and show it contains a \textit{``multi-frequency''} structure. 
    In a similar manner, we introduce the truncated MLE for the joint estimation problem with continuous phase angles in $[0,2\pi)$. 
    
    \item Inspired by~\cite{fan2021spectral} and the \textit{``multi-frequency''} nature of the MLE formulation, we propose a spectral method based on the novel MF-CPQR factorization. The MF-CPQR factorization is adjusted from the CPQR factorization to cope with information across multiple frequencies. Similarly, we also introduce an iterative MF-GPM by carefully designing steps of leveraging the \textit{``multi-frequency''} structure.
    
    \item We compare the performance of our proposed methods to state-of-the-art methods~\cite{fan2021spectral,chen2021non} on both discrete and continuous phase angles in $[0,2\pi)$ via a series of numerical experiments. Our proposed methods significantly outperform them in exact recovery of the cluster structure and error of phase synchronization. 
\end{itemize}

\subsection{Organization}
The rest of this paper is organized as follows. The formal definition of SBM-Ph, the MLE formulation of the joint estimation problem, and the extension to continuous phase angles, are detailed in Section~\ref{sec:Problem_Formulation}. Section~\ref{sec:MFCPQR} and~\ref{sec:MFGPM} present the spectral method based on the MF-CPQR factorization and the iterative MF-GPM, respectively. Numerical experiments are in Section~\ref{sec:exp}. Finally, we conclude the paper in Section~\ref{sec:conclusion}.

\subsection{Notations} Throughout this paper, we use $[n]$ to denote the set $\{1,2,\ldots,n\}$, and $\indfunc\{\cdot\}$ to denote the indicator function. The uppercase and lowercase letters in boldface are used to represent matrices and vectors, while normal letters are reserved for scalars. $\|\bm{X}\|_\text{F}$ and $\text{Tr}(\bm{X})$ denote the Frobenius norm and the trace of matrix $\bm{X}$, and $\|\bm{v}\|_2$ denotes the $\ell_2$ norm of the vector $\bm{v}$. The transpose and conjugate transpose of a matrix $\bm{X}$ (resp. a vector $\bm{x}$) are denoted by $\bm{X}^\top$ and $\bm{X}\hermconj$ (resp. $\bm{x}^\top$ and $\bm{x}\hermconj$), respectively. An $m\times n$ matrix of all zeros is denoted by $\bm{0}_{m\times n}$ (or $\bm{0}$, for brevity). An identity matrix of size $n\times n$ is defined as $\bm{I}_n$.  The complex conjugate of $x$ is denoted by $\overline{x}$. The inner product $\langle\cdot,\cdot\rangle$ between two scalars, vectors, and matrices are $\langle x, y\rangle = \overline{x}y$, $\langle \bm{x}, \bm{y}\rangle = \bm{x}\hermconj\bm{y}$, and $\langle \bm{X}, \bm{Y}\rangle = \text{Tr}(\bm{X}\hermconj\bm{Y})$, respectively. In terms of indexing, $(i,j)$th entry of $\bm{X}$ is denoted by $\bm{X}_{ij}$, and $i$th entry of $\bm{x}$ is denoted by $\bm{x}_i$. $\bm{X}_{i,\cdot}$ (resp. $\bm{X}_{\cdot,j}$) is used to denote $i$th row (resp. $j$th column) of $\bm{X}$. We use $\bm{X}_{i,j:}$ (resp. $\bm{X}_{i:,j}$) to denote the segment of the $i$th row (resp. $j$th column) from the $j$th entry (resp. $i$th entry) to the end, and $\bm{x}_{i:}$ to denote the segment from $i$th entry to the end. In addition, the sub-matrix of $\bm{X}$ from the $i$th row and $j$th column to the end is denoted by $\bm{X}_{i:,j:}$. Lastly, we use $\mO$ and $\Theta$ to denote the usual Big-O and Big-Theta notations. The notations are summarized in Table~\ref{table:notation}.

\begin{table}[h!]
	\caption{Notation table.}
	\centering
	\begin{tabular}{|c|c|}
		\hline
		[n] & Set of first $n$ positive integers: $1,\ldots, n$.\\ \hline
		$\indfunc\{\cdot\}$ & Indicator function. \\ \hline
		$\bm{X}$, $\bm{x}$, $x$ & Matrix, vector, scalar. \\ \hline
		$\bm{X}^\top$, $\bm{X}\hermconj$ & Transpose, conjugate transpose. \\ \hline
		$\overline{x}$ & Complex conjugate. \\ \hline
		$\left\langle\cdot,\cdot\right\rangle$ & Inner product. \\ \hline
		$\|\cdot\|_\mathrm{F}$ & Frobenius norm of a matrix. \\ \hline
		$\|\cdot\|_2$ & $\ell_2$ norm of a vector. \\ \hline
		$\bm{0}_{m\times n}$ (or $\bm{0}$) & All zero matrix of size $m\times n$. \\ \hline
		$\bm{I}_n$ & Identity matrix of size $n\times n$. \\ \hline
		$\bm{X}_{ij}$ & the $(i,j)$th entry of $\bm{X}$. \\ \hline
		$\bm{x}_{i}$ & the $i$th entry of $\bm{x}$. \\ \hline
		$\bm{X}_{i,\cdot}$ ($\bm{X}_{\cdot,j}$) & the $i$th row (the $j$th column) of $\bm{X}$. \\ \hline
		$\bm{X}_{i,j:}$ ($\bm{X}_{i:,j})$ & \tabincell{c}{Segment of the $i$th row (the $j$th column) from the $j$th \\  entry (the $i$th entry) to the end of the row (the column).}\\ \hline
		$\bm{x}_{i:}$ & Segment of the vector $\bm{x}$ from the $i$th entry to the end. \\ \hline
		$\bm{X}_{i:,j:}$ & \tabincell{c}{Sub-matrix of $\bm{X}$ from the $i$th row \\ and the $j$th column to the end.}\\ \hline
		$\mO$ & Big-O notation. \\ \hline
		$\Theta$ & Big-Theta notation. \\ \hline
	\end{tabular}
	\label{table:notation}
\end{table}

\section{Preliminaries}
\label{sec:prelim}

In this section, we introduce some important definitions that will be used in our algorithms later.

\begin{definition}[QR factorization]
\label{def:qr}
Given $\bm{X}\in\C^{m\times n}$, a QR factorization of $\bm{X}$ satisfies
\begin{equation*}
    \bm{X} = \bm{Q}\bm{R},
\end{equation*}
where $\bm{Q}\in\C^{m\times m}$ is a unitary matrix, and $\bm{R}\in\C^{m\times n}$ is an upper triangular matrix.
\end{definition}
Such factorization always exists for any $\bm{X}$. The most common methods for computing the QR factorization are Gram-Schmidt process~\cite{trefethen1997numerical}, and Householder transformation~\cite{bulirsch1991introduction}.

\begin{definition}[Column-pivoted QR factorization] 
\label{def:cpqr}
Let $\bm{X}\in\C^{m\times n}$ with $m \leq n$ has rank $m$. The column-pivoted QR factorization of $\bm{X}$ is the factorization
\begin{equation*}
    \bm{X}\bm{\Pi}_n=\bm{Q}\left[ \bm{R}_1, \, \bm{R}_2 \right],
\end{equation*}
as computed via the Golub-Businger algorithm~\cite{businger1971linear} where $\bm{\Pi}_n\in\{0,1\}^{n\times n}$ is a permutation matrix, $\bm{Q}$ is a unitary matrix, $\bm{R}_1$ is an upper triangular matrix, and $\bm{R}_2 \in \mathbb{C}^{m \times (n - m)}$.
\end{definition}
The ordinary QR factorization is proceeded on $\bm{X}$ from the first column to the last column in order, whereas the order of the CPQR factorization is indicated by $\bm{\Pi}_n$. 
We refer to~\cite{businger1971linear} for more details on the CPQR factorization. 

\begin{definition}[Projection onto $\mH$ in~\eqref{eq:SBM_MLE}]
    \label{def:proj_H}
For an arbitrary matrix $\bm{X}\in\R^{m\times n}$, we define
\begin{equation*}
    \mathcal{P}_\mH(\bm{X}):=\argmin_{\bm{H}\in\mH}\|\bm{H}-\bm{X}\|_\mathrm{F}=\argmax_{\bm{H}\in\mH}\langle\bm{H},\bm{X}\rangle
\end{equation*}
as the projection of $\bm{X}$ onto $\mH$.
\end{definition}
The projection aims to find the cluster structure that has the largest overall score given by $\bm{X}$. It is shown in \cite{wang2021optimal} that projection onto $\mH$ is equivalent to a \textit{minimum-cost assignment problem} (MCAP), and can be efficiently solved by the ``incremental algorithm" for MCAP~\cite[Section 3]{tokuyama1995geometric} with $\mO(n^2m\log{m})$ computational complexity. The uniqueness condition of the projection $\mathcal{P}_\mH(\bm{X})$ can be found in the proof of~\cite[Theorem 2.1]{tokuyama1995geometric} and~\cite[Theorem 2]{numata1993splitting}. If the solution is not unique, the ``incremental algorithm" for MCAP~\cite[Section 3]{tokuyama1995geometric} will generates a feasible projection randomly.

\section{Problem Formulation}
\label{sec:Problem_Formulation}
In this section, we formally define the probabilistic model, SBM-Ph, studied in this paper. We first consider discrete phase angles and formulate the corresponding MLE problem, which exhibits a \emph{multi-frequency} structure. Then, we extend the problem to continuous phase angles and formulate a truncated MLE problem.

\subsection{Stochastic Block Model with Discrete Relative Phase Angles}
\label{subsec:SBM_Ph}
SBM-Ph is considered in a network with $N$ nodes and $M\ge 2$ underlying clusters of equal size $s = \sfrac{N}{M}$. We assume each node $i\in[N]$ falls into one of $M$ underlying clusters with the assignment $\mM^*(i)\in [M]$, and is associated with an unknown phase angle $\theta^*_i\in\Omega$, where $\Omega:=\left\{0,\ldots,(2K_{\text{max}}+1)\Delta\right\}$ is a discretization of $[0,2\pi)$ with $\Delta = \sfrac{2\pi}{(2K_{\text{max}}+1)}$. We use $\mS^*_m$ to denote the set of nodes belonging to the $m$th cluster for all $m\in[M]$. 

SBM-Ph generates a random graph $\mG=([N], \mE)$ with the node set $[N]$ and the edge set $\mE\subseteq [N]\times [N]$. Each pair of nodes $(i,j)$ are connected independently with probability $p$ if $i$ and $j$ belong to the same cluster, or equivalently, $\mM^*(i)=\mM^*(j)$. Otherwise, $i$ and $j$ are connected independently with probability $q$ if $\mM^*(i)\neq\mM^*(j)$. Meanwhile, a relative phase angle $\theta_{ij}\in\Omega$ is observed on each edge $(i,j)\in\mE$. When $\mM^*(i)=\mM^*(j)$, we obtain $\theta_{ij} := (\theta^*_i - \theta^*_j) \mod 2\pi$. Otherwise, we observe $\theta_{ij}:=u_{ij}\sim\text{Unif}(\Omega)$, which is drawn uniformly at random from $\Omega$.

Our observation model can be represented by the \textit{observation matrix} $\bm{A}\in\C^{N\times N}$, which is a Hermitian matrix whose $(i,j)$th entry for any $i<j$ satisfies,
\begin{align}
    \label{eq:observation_model}
    \bm{A}_{ij}=
    \begin{cases}
    e^{\iota (\theta^*_i - \theta^*_j)}, & \textup{with prob } p \textup{ if } \mM^*(i) = \mM^*(j),\\
    e^{\iota u_{ij}}, &\textup{with prob } q \textup{ if } \mM^*(i) \neq \mM^*(j),\\
    0, &\textup{o.w.},
    \end{cases}
\end{align}
where $\bm{A}_{ji} = \overline{\bm{A}_{ij}}$. We also set the diagonal entry $\bm{A}_{ii} = 0, \forall i\in[N]$. Notice that a realization generated by the above observation matrix~\eqref{eq:observation_model} is a noisy version of the \textit{clean observation matrix} $\bm{A}^{\text{clean}}\in\C^{N\times N}$, whose $(i,j)$th entry satisfies,
\begin{align}
    \label{eq:clean_ob}
    \bm{A}^{\text{clean}}_{ij}=
    \begin{cases}
    e^{\iota (\theta^*_i - \theta^*_j)}, &\text{if }\mM^*(i) = \mM^*(j),\\
    0, &\textup{otherwise}.
    \end{cases} 
\end{align}
Specially, $\bm{A}$ is equal to $\bm{A}^{\text{clean}}$ when $p=1$ and $q=0$.

\begin{remark}
Unlike the observation matrix (or adjacency matrix) $\bm{A}_{\text{SBM}}$ in SBM~\cite{abbe2015community,abbe2015exact, abbe2017community, bandeira2018random} with only $\{0,1\}$-valued entries, $\bm{A}$ in~\eqref{eq:observation_model} extends to incorporating the relative phase angles $\theta_{ij}$ into edges. On the other hand, while entries of the observation matrix $\bm{A}_{\text{Ph}}$ in the phase synchronization problem~\cite{singer2011angular,perry2018message,gao2019multi} encode the the pairwise transformation information, they do not have the underlying $M$-cluster structure.
\end{remark}

\subsection{MLE with Multi-Frequency Nature}
\label{subsec:MLE}
Based on the observation matrix $\bm{A}$, we detail the MLE formulation for recovering the cluster structure and phase angles in this section. Given parameters, phase angles associated with nodes $\{\theta_i\in\Omega\}_{i=1}^N$ and the cluster structure $\{\mS_m\}_{m=1}^M$ of equal size $s$, the probability model of observing $\bm{A}_{ij}$ between node pair $(i,j)$ is 
\begin{equation*}
\begin{aligned}
&\Prob\left(\bm{A}_{ij}\left| \{\theta_i\in\Omega\}_{i=1}^N, \{\mS_m\}_{m=1}^M\right.\right)  \\
&=\begin{cases}
    p, &\text{if }\bm{A}_{ij} = e^{\iota(\theta_i-\theta_j)}\text{ and } \mM(i)=\mM(j),\\
    0, &\text{if }\bm{A}_{ij} \neq e^{\iota(\theta_i-\theta_j)}\text{ and } \mM(i)=\mM(j),\\
    1-p, &\text{if }\bm{A}_{ij}=0\text{ and } \mM(i)=\mM(j),\\
    \sfrac{q}{K}, &\text{if } \bm{A}_{ij} = e^{\iota u_{ij}} \text{ and } \mM(i)\neq \mM(j),\\
    1-q, &\text{if } \bm{A}_{ij}=0 \text{ and }\mM(i)\neq \mM(j),
\end{cases} 
\end{aligned}
\end{equation*}
where $\mM(\cdot)$ is the assignment function corresponding to the cluster structure $\{\mS_m\}_{m=1}^M$, and $K = 2 K_{\text{max}} + 1$. The likelihood function given observations on the edge set $\mE$ is
\begin{equation}
\label{eq:likelihood_func}
\begin{aligned}
&\Prob\left(\{\bm{A}_{ij}\}_{(i,j)\in\mE}\left| \{\theta_i\in\Omega\}_{i=1}^N, \{\mS_m\}_{m=1}^M\right.\right)\\
&=\prod_{\substack{\mM(i)=\mM(j)\\ (i,j)\in\mE}}p^{\indfunc\{\bm{A}_{ij} = e^{\iota(\theta_i-\theta_j)}\}} \prod_{\substack{\mM(i)\neq \mM(j) \\ (i,j)\in\mE}}\sfrac{q}{K}\\
&\times\prod_{\substack{\mM(i)=\mM(j)\\ (i,j)\in\mE}}0^{\indfunc\{\bm{A}_{ij} \neq e^{\iota(\theta_i-\theta_j)}\}},
\end{aligned}
\end{equation}
due to the independence among edges within $\mE$. Notice that maximizing the likelihood function~\eqref{eq:likelihood_func} is equal to maximizing the following log-likelihood function
\begin{equation}
\label{eq:log_likelihood}
\begin{aligned}
    &\log{ \Prob\left(\{\bm{A}_{ij}\}_{(i,j)\in\mE}\left| \{\theta_i\in\Omega\}_{i=1}^N, \{\mS_m\}_{m=1}^M\right.\right)} = \\
    &\sum_{\substack{\mM(i)=\mM(j)\\ (i,j)\in\mE}}\indfunc\{\bm{A}_{ij} = e^{\iota(\theta_i-\theta_j)}\}\log{p}+\sum_{\substack{\mM(i)\neq \mM(j)\\ (i,j)\in\mE}} \log{\sfrac{q}{K}}\\
    &+\sum_{\substack{\mM(i)=\mM(j)\\ (i,j)\in\mE}}\indfunc\{\bm{A}_{ij} \neq e^{\iota(\theta_i-\theta_j)}\}\log{0}.
\end{aligned}
\end{equation}
Given $0 < \sfrac{q}{K} < p$, maximizing~\eqref{eq:log_likelihood} is equivalent to
\begin{equation}
\label{eq:cost1}
    \max_{\substack{\{\theta_i\in\Omega\}_{i=1}^N\\ \{\mS_m\}_{m=1}^M}}\sum_{\substack{\mM(i)=\mM(j)\\ (i,j)\in\mE}}\indfunc\{\theta_{ij}=[(\theta_i-\theta_j) \mod 2\pi]\},
\end{equation}
by assuming $0\log{0}=0$ in~\eqref{eq:log_likelihood}.
By taking the FFT w.r.t. the support $\Omega$ of $((\theta_i-\theta_j) \mod 2\pi)$s and inverse FFT (IFFT) back,~\eqref{eq:cost1} is equivalent to
\begin{equation}
\label{eq:cost2}
    \max_{\substack{\{\theta_i\in\Omega\}_{i=1}^N\\ \{\mS_m\}_{m=1}^M}} \quad \sum_{k=-K_\text{max}}^{K_\text{max}}\sum_{m=1}^M\sum_{i,j\in \mS_m}\left\langle  \bm{A}^{(k)}_{ij}, e^{\iota k(\theta_i-\theta_j)}\right\rangle,
\end{equation}
where $\bm{A}^{(k)}$ is the $k$th entry-wise power of $\bm{A}$ with $\bm{A}^{(k)}_{ij} = e^{\iota k\theta_{ij}}$. 

As indicated by~\eqref{eq:cost2}, the MLE exhibits a \textit{multi-frequency} nature, where the $k$th frequency component is $\sum_{m=1}^M\sum_{i,j\in \mS_m}\langle  \bm{A}^{(1)}_{ij}, e^{\iota (\theta_i-\theta_j)}\rangle$ in~\eqref{eq:cost2}. Although the following program using the first frequency component 
\begin{equation}
\label{eq:single_cost}
    \max_{\substack{\{\theta_i\in\Omega\}_{i=1}^N\\ \{\mS_m\}_{m=1}^M}}\quad \sum_{m=1}^M\sum_{i,j\in \mS_m}\left\langle \bm{A}^{(1)}_{ij}, e^{\iota(\theta_i-\theta_j)} \right\rangle,
\end{equation}
is a reasonable formulation for the joint estimation problem as suggested by~\cite{fan2022joint,fan2021spectral,chen2021non}, it is indeed not a MLE formulation. One can show that~\eqref{eq:single_cost} is equivalent to
\begin{equation*}
    \max_{\substack{\{\theta_i\in\Omega\}_{i=1}^N\\ \{\mS_m\}_{m=1}^M}}\quad\sum_{\substack{\mM(i)=\mM(j)\\ (i,j)\in\mE}}\cos(\theta_{ij}-(\theta_i-\theta_j)),
\end{equation*}
which is not the MLE~\eqref{eq:cost1} of the joint estimation problem.

To proceed, we perform a change of optimization variables for~\eqref{eq:cost2}. By defining a unitary matrix $\bm{V}\in \C^{N\times M}$ whose $(i,m)$th entry satisfies
\begin{align}
\label{eq:V_form}
    \bm{V}_{im}:=
    \begin{cases}
    \frac{1}{\sqrt{s}}e^{\iota \theta_i}, &\text{if }i\in \mS_m (\text{or }\mM(i)=m),\\
    0, &\text{otherwise},
    \end{cases}
\end{align}
the cluster structure $\{\mS_m\}_{m=1}^M$ and the associated phase angles  $\{\theta_i\in\Omega\}_{i=1}^N$ are encoded into one simple unitary matrix $\bm{V}$. Then, the optimization program~\eqref{eq:cost2} can be reformulated as
\begin{equation}
\label{eq:cost4}
    \begin{aligned}
    \max_{\bm{V}\in \C^{N\times M}}\quad &\sum_{k=-K_\text{max}}^{K_\text{max}}\left\langle \bm{A}^{(k)},  \bm{V}^{(k)}\left(\bm{V}^{(k)}\right)\hermconj\right\rangle\\
    \text{s.t.} \qquad&\bm{V} \text{ satisfies the form~\eqref{eq:V_form}},
    \end{aligned}
\end{equation}
where each $\bm{V}^{(k)}$ is generated by $\bm{V}$ through the entry-wise power that satisfies
\begin{align}
\label{eq:Vk_form}
    \bm{V}^{(k)}_{im}:=
    \begin{cases}
    \frac{1}{\sqrt{s}}e^{\iota k\theta_i}, &\text{if }i\in \mS_m (\text{or }\mM(i)=m),\\
    0, &\text{otherwise}.
    \end{cases}
\end{align}

The optimization program~\eqref{eq:cost4} is non-convex, and is thus computationally intractable to be solved exactly. Although one can try SDP based approaches similar to~\cite{fan2022joint}, it is not guaranteed to obtain exact solutions to the MLE, let alone the high computational complexity when $N$ and $K_\text{max}$ are large. Therefore, we propose a spectral method based on the MF-CPQR factorization and an iterative MF-GPM in Section~\ref{sec:MFCPQR} and Section~\ref{sec:MFGPM}, respectively.

\subsection{Extension to Continuous Phase Angles: A Truncated MLE}
\label{subsec:ext}

We consider the joint estimation problem on a discretization of $[0,2\pi)$ in Section~\ref{subsec:SBM_Ph}, and then derive the MLE formulation in Section~\ref{subsec:MLE}. Now, we turn to the joint estimation problem with continuous phase angles in $[0,2\pi)$ ($\theta_i\in[0,2\pi),\forall i\in[N]$).

Following the similar steps as~\eqref{eq:likelihood_func}, \eqref{eq:log_likelihood}, \eqref{eq:cost1}, the MLE formulation is
\begin{equation}
\label{eq:MLE2}
    \max_{\substack{\{\theta_i\in[0,2\pi)\}_{i=1}^N\\ \{\mS_m\}_{m=1}^M}}\sum_{\substack{\mM(i)=\mM(j)\\ (i,j)\in\mE}}\indfunc([(\theta_i-\theta_j) \mod 2\pi]=\theta_{ij}).
\end{equation}
The MLE formulation~\eqref{eq:MLE2} is essentially equal to counting the times that $\delta([(\theta_i-\theta_j) \mod 2\pi]=\theta_{ij}) = \infty$, where $\delta(\cdot)$ is the Dirac delta function. We can express the Dirac delta function with its Fourier series expansion,
\begin{equation}
\label{eq:FSE}
\begin{aligned}
&\delta([(\theta_i-\theta_j) \mod 2\pi]=\theta_{ij}) =  \sum_{k=-\infty}^{+\infty}e^{\iota k(\theta_i-\theta_j)}e^{-\iota k\theta_{ij}} \\
&\approx \sum_{k=-K_\text{max}}^{K_\text{max}}e^{\iota k(\theta_i-\theta_j)}e^{-\iota k\theta_{ij}}.
\end{aligned}
\end{equation}
The straightforward truncation in~\eqref{eq:FSE} corresponds to approximating the Dirac delta with the \emph{Dirichlet kernel}. By this truncation, the problem in~\eqref{eq:MLE2} is converted to
\begin{equation}
   \label{eq:cont_cost3}
        \max_{\substack{\{\theta_i\in[0,2\pi)\}_{i=1}^N\\ \{\mS_m\}_{m=1}^M}}  \sum_{k=-K_\text{max}}^{K_\text{max}}\sum_{m=1}^M\sum_{i,j\in \mS_m}\left\langle  \bm{A}^{(k)}_{ij}, e^{\iota k(\theta_i-\theta_j)}\right\rangle. 
\end{equation}
The optimization program~\eqref{eq:cont_cost3} is a truncated MLE of the joint estimation problem with continuous phase angles of~\eqref{eq:MLE2}.

As one can observe from~\eqref{eq:cost2} and~\eqref{eq:cont_cost3}, the only difference is that $\theta_i\in\Omega$ is discrete in~\eqref{eq:cost2}, and $\theta_i\in [0,2\pi)$ is continuous in~\eqref{eq:cont_cost3}. Algorithms in Section~\ref{sec:MFCPQR} and~\ref{sec:MFGPM} can also be directly applied to the joint estimation problem with continuous phase angles after simple modification. Due to the similarity between the joint estimation problem and its continuous extension, we will only focus on the joint estimation problem on $\Omega$ (despite numerical experiments) in remaining parts of this paper for brevity.

\section{Spectral Method Based on the MF-CPQR Factorization}
\label{sec:MFCPQR}

In this section, we propose a spectral method based on the novel MF-CPQR factorization for the joint estimation problem. We start with introducing main steps and motivations of Algorithm~\ref{alg:spec_alg} in Section~\ref{subsec:motivation}. Section~\ref{subsec:MF-CPQR} states the novel algorithm, the MF-CPQR factorization, designed for our spectral method, together with the difference between the MF-CPQR factorization and the CPQR factorization. In Section~\ref{subsec:spec_complexity}, we discuss the computational complexity of our proposed algorithm in details.

Our spectral method based on the MF-CPQR factorization is inspired by the CPQR-type algorithms~\cite{damle2019simple,fan2021spectral}, together with the \textit{multi-frequency} nature of the MLE formulation~\eqref{eq:cost4}. Similar to the CPQR-type algorithms, Algorithm~\ref{alg:spec_alg} is deterministic and free of any initialization. Meanwhile, in terms of computational complexity, Algorithm~\ref{alg:spec_alg} scales linearly w.r.t. the number of edges $|\mE|$ and near-linearly w.r.t. $K_\text{max}$.

\begin{algorithm}[t!]
\footnotesize
\SetAlgoLined
\SetNoFillComment
\setcounter{AlgoLine}{0}
\KwIn{The observation matrix $\bm{A}$, and the number of clusters $M$.}

(Eigendecomposition) For $k=-K_\text{max},\ldots, K_\text{max}$, compute the top $M$ eigenvectors $\bm{\Phi}^{(k)} \in \mathbb{C}^{N \times M}$ of $\bm{A}^{(k)}$ such that $\left(\bm{\Phi}^{(k)}\right)\hermconj \bm{\Phi}^{(k)} = \bm{I}_{M}$

(MF-CPQR factorization) Compute the multi-frequency column-pivoted QR factorization (detailed in Algorithm~\ref{alg:CPQR}) of $\left\{\left(\bm{\Phi}^{(k)}\right)^\top\right\}_{k=-K_\text{max}}^{K_\text{max}}$, which yields
    \begin{equation}
    \left(\bm{\Phi}^{(k)}\right)^\top\bm{\Pi}_N = \bm{Q}^{(k)}\bm{R}^{(k)}\Rightarrow \left(\bm{\Phi}^{(k)}\right)^\top=\bm{Q}^{(k)}\bm{R}^{(k)}\bm{\Pi}_N^\top
    \end{equation}
    Update $\bm{R}^{(k)} \leftarrow \bm{R}^{(k)}\bm{\Pi}_N^\top, \forall k=-K_\text{max},\ldots, K_\text{max}$
    
(Recovery of the cluster structure and the phase angles)
    For each node $i\in[N]$, assign its cluster as
    \begin{equation}
    \label{eq:cluster1}
        \hat{\mM}(i)\leftarrow \argmax_{m\in[M]}\quad\left\{\max_{\theta_i\in\Omega}\sum_{k=-K_\text{max}}^{K_\text{max}}\left\langle  e^{\iota k\theta_i}, \bm{R}^{(k)}_{mi}\right\rangle\right\}
    \end{equation}
    Then estimate the phase angle given the recovered cluster assignment $\hat{\mM}(i)$
    \begin{equation}
    \label{eq:phase1}
        \hat{\theta}_i \leftarrow \argmax_{\theta_i\in\Omega} \sum_{k=-K_\text{max}}^{K_\text{max}}\left\langle e^{\iota k\theta_i}, \bm{R}^{(k)}_{\hat{\mM}(i)i}\right\rangle
    \end{equation}

\KwOut{\textup{Estimated cluster structure $\{\hat{\mM}(i)\}_{i = 1}^N$ and estimated phase angles $\{\hat{\theta}_i\}_{i = 1}^N$}}
\caption{The spectral method based on the MF-CPQR factorization}
\label{alg:spec_alg}
\end{algorithm}
\vspace{-0.3cm}
\subsection{Motivations}
\label{subsec:motivation}

Algorithm~\ref{alg:spec_alg} consists of three steps: i) Eigendecomposition of $\bm{A}^{(k)}$, ii) MF-CPQR factorization, and iii) Recovery of the cluster structure and phase angles. It first computes matrices $\{\bm{\Phi}^{(k)}\}_{k=-K_\text{max}}^{K_\text{max}}$ that contain the top $M$ eigenvectors of each $\bm{A}^{(k)}$ via eigendecomposition. Secondly, matrices $\{\bm{R}^{(k)}\}_{k=-K_\text{max}}^{K_\text{max}}$ are obtained through the MF-CPQR factorization, which is detailed in Algorithm~\ref{alg:CPQR}. The last step is recovering the cluster structure and associated phase angles based on $\{\bm{R}^{(k)}\}_{k=-K_\text{max}}^{K_\text{max}}$ via~\eqref{eq:cluster1} and~\eqref{eq:phase1}.

In terms of motivations for Algorithm~\ref{alg:spec_alg}, we start from the MLE formulation~\eqref{eq:cost4}. We first relax~\eqref{eq:cost4} by replacing the constraints in~\eqref{eq:V_form} with $\bm{V}\hermconj\bm{V}=\bm{I}_M$,
\begin{equation}
\label{eq:releax_cost}
\begin{aligned}
    \bm{\Phi}=&\argmax_{\bm{V}\in\C^{N\times M}}\quad \sum_{k=-K_\text{max}}^{K_\text{max}}\left\langle \bm{A}^{(k)}, \bm{V}^{(k)}\left(\bm{V}^{(k)}\right)\hermconj\right\rangle\\
    &\text{s.t. }\bm{V}\hermconj\bm{V}=\bm{I}_M,
\end{aligned}
\end{equation}
by noticing that $\bm{V}$ in~\eqref{eq:V_form} forms an orthonormal basis. The optimization problem in~\eqref{eq:releax_cost} is still non-convex and there is no simple spectral method that can directly solve the problem. One approach is to relax the dependency of $\bm{V}^{(k)}$ among different frequencies and split~\eqref{eq:releax_cost} into different frequencies, and that is, for $k=-K_\text{max},\ldots, K_\text{max}$, we have
\vspace{-0.2cm}
\begin{equation}
\label{eq:releax_cost2}
\begin{aligned}
\bm{\Phi}^{(k)}=&\argmax_{\bm{V}^{(k)}\in\C^{N\times M}}\quad \left\langle\bm{A}^{(k)}, \bm{V}^{(k)}\left(\bm{V}^{(k)}\right)\hermconj\right\rangle\\ &\text{s.t. }\left(\bm{V}^{(k)}\right)\hermconj\bm{V}^{(k)}=\bm{I}_M.
\end{aligned}
\end{equation}
The optimizer of~\eqref{eq:releax_cost2} is the matrix that contains the top $M$ eigenvectors of $\bm{A}^{(k)}$ denoted by $\bm{\Phi}^{(k)}\in\C^{N\times M}$. This accounts for step 1 (eigendecomposition) in Algorithm~\ref{alg:spec_alg}.

In fact, one can infer the cluster structure from $\{\bm{\Phi}^{(k)}\}_{k=-K_\text{max}}^{K_\text{max}}$. To see this, for $k=-K_\text{max},\ldots, K_\text{max}$, we split $\bm{A}^{(k)}$ into deterministic and random parts:
\begin{equation}
\label{eq:decomp}
    \bm{A}^{(k)}=\E[\bm{A}^{(k)}]+(\bm{A}^{(k)}-\E[\bm{A}^{(k)}]) = \E[\bm{A}^{(k)}] + \bm{\Delta}^{(k)},
\end{equation}
where $\E[\bm{A}^{(k)}]=p\bm{A}^{(k)}_{\text{clean}}$ with $\bm{A}^{(k)}_{\text{clean}}$ being the entry-wise $k$th power of $\bm{A}_{\text{clean}}$~\eqref{eq:clean_ob}, and the residual $\bm{\Delta}^{(k)}$ is a random perturbation with $\E[\bm{\Delta}^{(k)}]=\bm{0}$. Obviously, each $\E[\bm{A}^{(k)}]$ is a low rank matrix that satisfies the following eigendecomposition:
\begin{equation*}
\begin{aligned}
&\E[\bm{A}^{(k)}] = ps\sum_{m=1}^M \bm{\Psi}^{(k)}_{\cdot,m}\left(\bm{\Psi}^{(k)}_{\cdot,m}\right)\hermconj,\\
&\text{with}\quad \bm{\Psi}^{(k)}_{im}:=
    \begin{cases}
    \frac{1}{\sqrt{s}}e^{\iota k\theta_i^*},\quad&\text{if } i\in\mS_m^*,\\
    0, &\text{otherwise},
    \end{cases}
\end{aligned}
\end{equation*}
where $\bm{\Psi}^{(k)}\in\C^{N\times M}$ is a matrix defined in a similar manner as $\bm{V}^{(k)}$ in~\eqref{eq:Vk_form}, and satisfies $\left(\bm{\Psi}^{(k)}\right)\hermconj\bm{\Psi}^{(k)}=\bm{I}_M$. Then, for $k=-K_\text{max},\ldots, K_\text{max}$ (except for $k=0$), the non-zero entry in each row of $\bm{\Psi}^{(k)}$ indicates the underlying cluster assignment $\mM^*(i)$ and the exact phase angle $\theta^*_i$ of node $i$.

Therefore, to recover the cluster structure and associated phase angles, it suffices to extract $\{\bm{\Psi}^{(k)}\}_{k=-K_\text{max}}^{K_\text{max}}$ from $\{\bm{\Phi}^{(k)}\}_{k=-K_\text{max}}^{K_\text{max}}$. For the ease of illustration, we first consider the case when $p=1$ and $q=0$. This indicates, for $k=-K_\text{max},\ldots, K_\text{max}$, $\bm{A}^{(k)} = \bm{A}^{(k)}_{\text{clean}}$, $\bm{\Delta}^{(k)}=\bm{0}$, and $\bm{\Phi}^{(k)}=\bm{\Psi}^{(k)}\bm{O}^{(k)}$, where $\bm{O}^{(k)}\in\C^{M\times M}$ is some unitary matrix. However, $\{\bm{O}^{(k)}\}_{k=-K_\text{max}}^{K_\text{max}}$ are unknown and even not synchronized among all frequencies. To address this issue, the MF-CPQR factorization is introduced. Here, we assume that the first $s$ nodes are from the cluster $\mS^*_1$, the following $s$ nodes are from $\mS^*_2$, and so on. Applying the MF-CPQR factorization (step 2) in Algorithm~\ref{alg:spec_alg} yields (assume $\bm{\Pi}_N=\bm{I}_N$)
\begin{equation}
    \begin{aligned}
    &\left(\bm{\Phi}^{(k)}\right)^\top = \left(\bm{O}^{(k)}\right)^\top \left(\bm{\Psi}^{(k)}\right)^\top= \frac{1}{\sqrt{s}}\left(\bm{O}^{(k)}\right)^\top\times\\
    &\begin{bmatrix}
    e^{\iota k\theta_1^*} &\cdots &e^{\iota k\theta_s^*} &\cdots & 0 &\cdots &0\\
    \vdots &\vdots &\vdots &\ddots &\vdots &\vdots &\vdots\\
    0 &\cdots &0 &\cdots &e^{\iota k\theta_{N-s+1}^*} &\cdots &e^{\iota k\theta_N^*}
    \end{bmatrix}\\
    &= 
    \underbrace{
    \left(\bm{O}^{(k)}\right)^\top 
    \begin{bmatrix}
    e^{\iota k\theta_1^*} &\cdots &0\\
    \vdots &\ddots &\vdots \\
    0 &\cdots & e^{\iota k\theta_{N-s+1}^*}
    \end{bmatrix}}_{=:\bm{Q}^{(k)}}\times\\
    &\underbrace{
    \begin{bmatrix}
    1&\cdots &e^{\iota k(\theta^*_s - \theta^*_1)} &\cdots &0 &\cdots &0\\
    \vdots &\vdots &\vdots &\ddots &\vdots &\vdots &\vdots\\
    0 &\cdots &0 &\cdots &1&\cdots &e^{\iota k(\theta^*_N - \theta^*_{N-s+1})}
    \end{bmatrix}}_{ =:\bm{R}^{(k)}}\\
    &= \bm{Q}^{(k)}\bm{R}^{(k)},
    \end{aligned}
    \label{eq:QR_general}
\end{equation}
for $k=-K_\text{max},\ldots, K_\text{max}$. Therefore, each $\bm{Q}^{(k)}\in\C^{M\times M}$ is a unitary matrix that includes the unknown unitary matrix $\bm{O}^{(k)}$, and each $\bm{R}^{(k)}\in \C^{M\times N}$ is a matrix excludes $\bm{O}^{(k)}$. More significantly, $\{\bm{R}^{(k)}\}_{k=-K_\text{max}}^{K_\text{max}}$ contains all the information needed to recover the cluster structure and associated phase angles.

To recover the cluster structure, the CPQR-type algorithm~\cite{fan2021spectral} only uses $\bm{R}^{(1)}$. By noticing that for each node $i$, the $i$th column of $\bm{R}^{(1)}$ (e.g., $\bm{R}^{(1)}_{\cdot,i}$) is sparse (its $m$th entry $\bm{R}^{(1)}_{mi}$ is nonzero if and only if $m=\mM^*(i)$), one can determine the cluster assignment of node $i$ by the position of the nonzero entry. Meanwhile, the associated phase angle can also be determined by obtaining the phase angle from the nonzero entry (up to some global phase transition in the same cluster). When the observation $\bm{A}$ is noisy, the CPQR-type algorithm recovers the cluster structure and associated phase angle of node $i$ by the position of the entry with the largest amplitude. The following Theorem~\ref{thm:1} proves as long as the perturbation to $\E[\bm{A}^{(k)}]$ is less than a certain threshold, $\bm{\Phi}^{(k)}$ is still close to $\bm{\Psi}^{(k)}\bm{O}^{(k)}$, for $k=-K_{\text{max}},\ldots,K_{\text{max}}$ (except for $k=0$).

\begin{theorem}[Row-wise error bound, adapted from~\cite{fan2021spectral}]
\label{thm:1}
Given a network with $N$ nodes and $M=2$ underlying clusters, for a sufficiently large $N$, we suppose
\begin{equation*}
    \eta:=\frac{\sqrt{(p(1-p)+q)\log{N}}}{p\sqrt{N}}\le c_0
\end{equation*}
for some small constant $c_0$. Consequently, with probability at least $1-\mO(N^{-1})$,
\begin{equation*}
    \max_{i\in[N]}\quad \left\|\bm{\Phi}^{(k)}_{i,\cdot}-\bm{\Psi}^{(k)}_{i,\cdot}\bm{O}^{(k)}\right\|_2\lesssim\frac{\eta}{\sqrt{N}},
\end{equation*}
where $\bm{O}^{(k)}=\mathcal{P}((\bm{\Psi}^{(k)})\hermconj\bm{\Phi}^{(k)})$.
\end{theorem}
Theorem~\ref{thm:1} guarantees that i) amplitudes of other entries are less than the entry indicating the true cluster structure with high probability, ii) the phase angle information is preserved with high fidelity. Theorem~\ref{thm:1} can be proved by following the same routines as~\cite{fan2021spectral} by replacing the orthogonal group element $\bm{O}_i$ with the $U(1)$ group element (e.g., $e^{\iota \theta_i}$). The reason why Theorem~\ref{thm:1} holds for $k=-K_{\text{max}},\ldots,K_{\text{max}}$ (despite $k=0$) is due to statistics of random perturbations $\{\bm{\Delta}^{(k)}\}_k$ in~\eqref{eq:decomp} do not change among different frequencies. This is because the noise models of $\bm{A}^{(k)}$ and $\bm{A}$ are the same. More specifically, the noisy entry $e^{\iota u_{ij}}:u_{ij}\sim\text{Unif}(\Omega)$ in~\eqref{eq:observation_model} has the same statistics as $e^{\iota k u_{ij}}$ in $\bm{A}^{(k)}$ due to the fact that $k u_{ij}$ still yields the distribution $\text{Unif}(\Omega)$. 

\begin{figure}[htbp]
    \centering
    \includegraphics[width = 0.95\columnwidth]{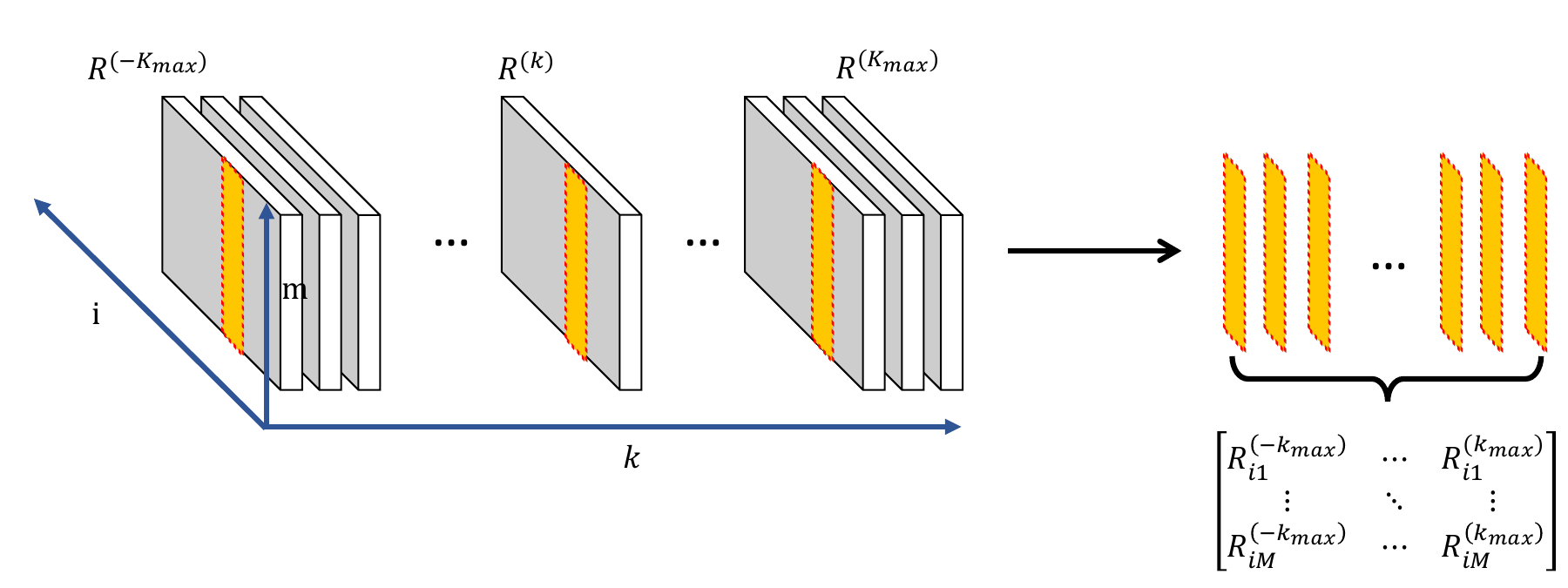}
    \caption{Illustration of step 3 in Algorithm~\ref{alg:spec_alg}. For each $i\in[N]$, all the $i$th columns in $\left\{\bm{R}^{(k)}\right\}_{k=-K_\textup{max}}^{K_\textup{max}}$ are extracted to estimate the cluster assignment and phase angle following~\eqref{eq:cluster1} and~\eqref{eq:phase1}.}
    \label{fig:R_k}
\end{figure}

Note that the CPQR-type algorithm in~\cite{fan2021spectral} is not developed from the MLE formulation~\eqref{eq:cost4} of the joint estimation problem, and thus does not capture the \textit{multi-frequency} nature. In this paper, we leverage  $\{\bm{R}^{(k)}\}_{k=-K_\text{max}}^{K_\text{max}}$ that contain information about the cluster structure and associated phase angles across multiple frequencies (step 3). Specifically, we first consider the same case as that in~\eqref{eq:QR_general} for intuition. As illustrated in Figure~\ref{fig:R_k}, the matrix concatenated by the $i$th ($i\le s$) columns across all frequencies is
\begin{equation*}
\begin{aligned}
    &\begin{bmatrix}
    \bm{R}_{i1}^{(-K_\textup{max})} &\cdots & \bm{R}_{i1}^{(k)}&\cdots &\bm{R}_{i1}^{(K_\textup{max})}\\
    \vdots &\vdots &\vdots &\ddots &\vdots\\
    \bm{R}_{iM}^{(-K_\textup{max})} &\cdots & \bm{R}_{iM}^{(k)}&\cdots &\bm{R}_{iM}^{(K_\textup{max})}
    \end{bmatrix} = \frac{1}{\sqrt{s}}\times\\
    &
    \begin{bmatrix}
   e^{-\iota K_\textup{max}(\theta_i^*-\theta_1^*)}&\cdots & e^{\iota k(\theta_i^*-\theta_1^*)}&\cdots &e^{\iota K_\textup{max}(\theta_i^*-\theta_1^*)}\\
   0&\cdots & 0&\cdots &0\\
    \vdots &\vdots &\vdots &\ddots &\vdots\\
    0 &\cdots & 0&\cdots &0
    \end{bmatrix}.
\end{aligned}
\end{equation*}
The cluster assignment of $i$ can be acquired by finding the non-sparse row of the above matrix, and the phase angle can be determined by evaluating the non-sparse row (e.g., FFT). When the observation $\bm{A}$ is noisy, \eqref{eq:cluster1} and~\eqref{eq:phase1} are used to estimate the cluster structure and associated phase angles, which can be interpreted as checking the consistency or conducting majority vote among all frequencies. The performance is expected to be at least as good as the CPQR-type algorithm. This is because each $\bm{\Phi}^{(k)}$ has the same theoretical guarantee as the CPQR-type algorithm according to Theorem~\ref{thm:1}, and \eqref{eq:cluster1} \eqref{eq:phase1} are just checking the consistency across all frequencies. In Section~\ref{sec:exp}, we will show that our proposed spectral method based on the MF-CPQR factorization is capable of significantly outperforming the CPQR-type algorithm. 

Besides, for the joint estimation problem with continuous phase angles, \eqref{eq:cluster1} and \eqref{eq:phase1} will be modified as  
\begin{align*}
&\hat{\mM}(i)\leftarrow \argmax_{m\in[M]}\quad\left\{\max_{\theta_i\in [0,2\pi)}\quad\sum_{k=-K_\text{max}}^{K_\text{max}}\left\langle  e^{\iota k\theta_i}, \bm{R}^{(k)}_{mi}\right\rangle\right\}, \\
&\hat{\mM}(i)\leftarrow \argmax_{m\in[M]}\quad\left\{\max_{\theta_i\in [0,2\pi)}\quad\sum_{k=-K_\text{max}}^{K_\text{max}}\left\langle  e^{\iota k\theta_i}, \bm{R}^{(k)}_{mi}\right\rangle\right\}, \\
&\hat{\theta}_i \leftarrow \argmax_{\theta_i\in [0,2\pi)}\quad \sum_{k=-K_\text{max}}^{K_\text{max}}\left\langle e^{\iota k\theta_i}, \bm{R}^{(k)}_{\hat{\mM}(i)i}\right\rangle.
\end{align*}
Solving the max problem over $[0,2\pi)$ is infeasible in general. Instead, one can apply the zero-padding and FFT for an approximate solution with any desired precision. Specifically, in estimating the cluster assignment, by padding zeros to $[\bm{R}^{(-K_\textup{max})}_{mi},\ldots,\bm{R}^{(k)}_{mi},\ldots,\bm{R}^{(K_\textup{max})}_{mi}]$ as $[0,\ldots,0,\bm{R}^{(-K_\textup{max})}_{mi},\ldots,\bm{R}^{(k)}_{mi},\ldots,\bm{R}^{(K_\textup{max})}_{mi},0,\ldots,0]$, taking the FFT, and finding the entry with largest real part, $(\arg)\max_{\theta_i\in [0,2\pi)}\sum_{k=-K_\text{max}}^{K_\text{max}}\left\langle  e^{\iota k\theta_i}, \bm{R}^{(k)}_{mi}\right\rangle$ can be solved approximately, where the precision is determined by the number of padded zeros.

\begin{algorithm}[t!]
\footnotesize
\DontPrintSemicolon
\SetAlgoLined
\SetNoFillComment

\KwIn{The set of eigenvectors $\left\{\left(\bm{\Phi}^{(k)}\right)\right\}_{k=-K_\text{max}}^{K_\text{max}}$}
\kwInit{$\bm{Q}^{(k)} \leftarrow \bm{I}_{M}$, $\bm{R}^{(k)} \leftarrow \left(\bm{\Phi}^{(k)}\right)^\top, \forall i=-K_\text{max},\ldots, K_\text{max}$, and $\bm{\Pi}_{N} \leftarrow \bm{I}_{N}$}

\For{$m = 1, 2, \ldots, M$}{ 
\tcc{Pivot selection}
\For{$j = m, m+1, \ldots, N$}{
Compute the residual $\rho_j \leftarrow \sum_{k=-K_\text{max}}^{K_\text{max}}\|\bm{R}^{(k)}_{m:, \; j}\|_2$
}
Determine the pivot $ j^* \leftarrow \argmax_{j = m,\ldots,N} \quad\rho_j$

For both $\{\bm{R}^{(k)}\}_{k=1}^{K-1}$ and $\bm{\Pi}_{N}$, swap the $m$th column with the pivot ($j^*$th) column

\tcc{One step QR factorization for all frequencies}

\For{$k = -K_\text{max}, \ldots, K_\text{max}$}{
Apply one step QR factorization in Algorithm~\ref{alg:Householder} on $\bm{R}_{m:,m:}^{(k)}$, and  get  $\widetilde{\bm{Q}}^{(k)}_{m:,m:}$ and $\widetilde{\bm{R}}^{(k)}_{m:,m:}$

 Update $\bm{Q}_m^{(k)} \leftarrow \begin{bmatrix}
 \bm{I}_{m-1} &\bm{0}\\
 \bm{0} &\widetilde{\bm{Q}}^{(k)}_{m:,m:}
 \end{bmatrix}$

Update $\bm{R}^{(k)}_{m:,m:} \leftarrow \widetilde{\bm{R}}^{(k)}_{m:,m:}$ and $\bm{Q}^{(k)} \leftarrow \bm{Q}^{(k)}\bm{Q}_m^{(k)}$

}
}
\KwOut{$\{\bm{Q}^{(k)}\}_{k=-K_\text{max}}^{K_\text{max}}$, $\{\bm{R}^{(k)}\}_{k=-K_\text{max}}^{K_\text{max}}$, and $\bm{\Pi}_{N}$} 
\caption{MF-CPQR factorization}
\label{alg:CPQR}
\end{algorithm}

\subsection{MF-CPQR Factorization}
\label{subsec:MF-CPQR}

\begin{algorithm}[t!]
\footnotesize
\DontPrintSemicolon
\SetAlgoLined
\SetNoFillComment
\KwIn{A matrix $\bm{X} \in \mathbb{C}^{n\times n}$}

\tcc{Householder transformation}

$\bm{r} \leftarrow \bm{X}_{\cdot,1}$  

$\theta \leftarrow -e^{\iota\angle \bm{r}_1}\|\bm{r}\|$, where $\angle \bm{r}_1$ is the phase angle of $\bm{r}_1$   

$\bm{u} \leftarrow \bm{r} - \theta\bm{e}$, where $\bm{e} = [1, 0, \ldots, 0]^\top$

$\bm{v} \leftarrow \sfrac{\bm{u}}{\|\bm{u}\|}$

$\bm{Q} \leftarrow \bm{I}_n - 2\bm{v}\bm{v}\hermconj$

$\bm{X}\leftarrow \bm{Q}\bm{X}$

$\bm{X}_{1,\cdot}\leftarrow e^{-\iota \angle \bm{X}_{11}} X_{1,\cdot}$ 

$\bm{Q}_{\cdot,1}\leftarrow e^{\iota \angle \bm{X}_{11}} \bm{Q}_{\cdot,1}$ 

\KwOut{$\bm{Q}$ and $\bm{R}$.}
\caption{One step QR factorization using Householder transformation}
\label{alg:Householder}
\end{algorithm}

As stated in Definition~\ref{def:cpqr}, the difference between the ordinary QR factorization and the CPQR factorization is selecting appropriate pivot ordering (encoded in $\bm{\Pi}_N$). The CPQR factorization attempts to find a subset of columns that are as most linearly independent as possible and are used to determine the basis. In this paper, the CPQR factorization across multiple frequencies is developed to cope with the \textit{multi-frequency} structure of the MLE formulation.

\begin{definition}[Multi-frequency column-pivoted QR factorization] 
\label{def:mfcpqr}
Let $\bm{X}^{(k)} \in\C^{m\times n}$ with $m \leq n$ has rank $m$ for $k = -K_\text{max}, \dots, K_\text{max}$. The multi-frequency column-pivoted QR factorization of $\bm{X}^{(k)}$ is the factorization
\begin{equation*}
    \bm{X}^{(k)} \bm{\Pi}_n= \bm{Q}^{(k)} \left[ \bm{R}^{(k)}_1, \, \bm{R}^{(k)}_2 \right],
\end{equation*}
as computed via Algorithm~\ref{alg:CPQR} where $\bm{\Pi}_n\in\{0,1\}^{n\times n}$ is a permutation matrix fixed for all $k = -K_\text{max}, \dots, K_\text{max}$, $\bm{Q}^{(k)}$ is a unitary matrix, $\bm{R}^{(k)}_1$ is an upper triangular matrix, and $\bm{R}^{(k)}_2 \in \mathbb{C}^{m \times (n - m)}$.
\end{definition}

It requires to i) obtain the same subset of columns among all frequencies that are as most linearly independent as possible, and ii) use the same pivot ordering (or $\bm{\Pi}_N$) among all frequencies. The former promotes the cluster structure estimation performance because each node $i$ (other than the pivots) is assigned to a cluster mainly according to the similarities between the column $i$ and the columns of pivots, the latter ensures the validity of~\eqref{eq:cluster1} and \eqref{eq:phase1}.

The MF-CPQR factorization is detailed in Algorithm~\ref{alg:CPQR}, where the Householder transform~\cite{bulirsch1991introduction} (Algorithm~\ref{alg:Householder}) is adopted for a better numerical stability. Specifically, the novel MF-CPQR factorization is different from the ordinary CPQR~\cite{golub1996matrix,trefethen1997numerical} in the pivot selection. The pivot is determined by finding the column with the largest summation of $\ell_2$ norm of residuals over all frequencies (see line 3 in Algorithm~\ref{alg:CPQR}).

\begin{table}[h!]
	\caption{The computational complexity of Algorithm~\ref{alg:spec_alg} in each step.}
	\centering
	\begin{tabular}{l||c|}
		\hline\hline
		Steps & Computational Complexity\\ \hline
		1. Eigendecomposition &  $\mO(K_\text{max}|\mE|)$     \\ \cline{2-2}
		2. MF-CPQR factorization & $\mO(K_\text{max}N)$ \\ \cline{2-2}
		3. Clustering by~\eqref{eq:cluster1} & $\mO(NK_\text{max}\log{K_\text{max}})$\\
		\cline{2-2}
		4. Phase synchronization by~\eqref{eq:phase1} & $\mO(N)$\\
		\hline
		Total complexity &  $\mO(K_\text{max}(|\mE|+N\log{K_\text{max}}))$\\
		\hline\hline
	\end{tabular}
	\label{table:comp1}
\end{table}

\subsection{Computational Complexity}
\label{subsec:spec_complexity}

In this section, the computational complexity of Algorithm~\ref{alg:spec_alg} is summarized step by step in Table~\ref{table:comp1}. Here, we suppose $M=\Theta(1)$. First, it consists of $\mO(K_\text{max})$ times of eigendecomposition for $M$ eigenvectors, which is $\mO(|\mE|)$ per time if using Lanczos method~\cite{stewart2002krylov}. For the MF-CPQR factorization, it consists of $M$ times of column pivoting ($\mO(NK_\text{max})$ per time) and $MK_\text{max}$ times of one step QR factorization ($\mO(N)$ per step). In terms of recovering the cluster structure, we first compute $MN$ times of FFT for length-$K_\text{max}$ vectors ($\mO(K_\text{max}\log{K_\text{max}})$ per vector) and then compute the maximums ($\mO(NK_\text{max})+\mO(N)$). Since the FFT of $\{\bm{R}^{(k)}\}_{k=-K_\text{max}}^{K_\text{max}}$ is already computed, it is only $\mO(N)$ for synchronizing the phase angles. Overall, the computational cost is linear with the number of edges $|\mE|$ and nearly linear in $K_\text{max}$. When the network $\mG$ is densely connected with $|\mE|=\mO(N^2)$, Algorithm~\ref{alg:spec_alg} ends up with $\mO(K_\text{max}N^2)$ if $\log{K_\text{max}}<N$. However, if $|\mE|=o(N^2)$, the complexity of Algorithm~\ref{alg:spec_alg} will be reduced. For instance, in the case when $|\mE|=\mO(N\log{N})$ or $|\mE|=\mO(N)$, which is very common as shown in~\cite{leskovec2008statistical}, the complexity of Algorithm~\ref{alg:spec_alg} will be $\mO(K_\text{max}N\max\{\log{N},\log{K_\text{max}}\})$ or $\mO(K_\text{max}N\log{K_\text{max}}\})$, respectively.

\section{Iterative Multi-Frequency Generalized Power Method}
\label{sec:MFGPM}

\begin{algorithm}[t!]
\footnotesize
\DontPrintSemicolon
\SetAlgoLined
\SetNoFillComment

\KwIn{The observation matrix $\bm{A}$, the initialization $\{\mS_m\}_{m=1}^M$ and $\{\theta_i\in\Omega\}_{i=1}^N$, and the number of iterations $T$}
Construct $\{\widehat{\bm{V}}^{(k),0}\}_{k=-K_\text{max}}^{K_\text{max}}$ using $\{\mS_m\}_{m=1}^M$ and $\{\theta_i\in\Omega\}_{i=1}^N$ according to~\eqref{eq:Vk_form}

\For{$t = 0, 1, \ldots, T-1$}{ 
\tcc{Matrix multiplication}
For $k=-K_\text{max}, \ldots, K_\text{max}$, compute the matrix multiplication $\widehat{\bm{V}}^{(k),t+1}\leftarrow \bm{A}^{(k)}\widehat{\bm{V}}^{(k),t}$

\tcc{Combine information across multiple frequencies}
Compute $\widehat{\bm{V}}^{\text{max}, t+1}\in\R^{N\times M}$, whose $(i,m)$th entry satisfies
\begin{equation}
\label{eq:v_hat}
    \widehat{\bm{V}}_{im}^{\text{max}, t+1} \leftarrow \max_{\theta_i\in\Omega}\quad\sum_{k=-K_\text{max}}^{K_\text{max}}\left\langle e^{\iota k\theta_i}, \widehat{\bm{V}}^{(k),t+1}_{im} \right\rangle
\end{equation}

\tcc{Recovery of the cluster structure and associated phase angles}
For each node $i\in[N]$, assign its cluster assignment as
\begin{equation*}
    \hat{\mM}(i)\leftarrow\argmax_{m\in[M]}\quad\widehat{\bm{H}}^{t+1}_{i,\cdot}, \text{ where }\widehat{\bm{H}}^{t+1}\leftarrow\mathcal{P}_{\mH}(\widehat{\bm{V}}^{\text{max}, t+1})
\end{equation*}
then estimate the associated phase angle given the estimated cluster assignment $\hat{\mM}(i)$
\begin{gather}
\label{eq:phase2}
    \hat{\theta}_i \leftarrow \argmax_{\theta_i\in\Omega} \sum_{k=-K_\text{max}}^{K_\text{max}}\left\langle e^{\iota k\theta_i}, \widehat{\bm{V}}^{(k),t+1}_{i\hat{\mM}(i)}\right\rangle
\end{gather}

Construct $\{\widehat{\bm{V}}^{(k),t+1}\}_{k=-K_\text{max}}^{K_\text{max}}$ using $\{\hat{\mM}(i)\}_{i = 1}^N$ and $\{\hat{\theta}_i\}_{i = 1}^N$ according to~\eqref{eq:Vk_form}
}

\KwOut{\textup{Estimated cluster structure $\{\hat{\mM}(i)\}_{i = 1}^N$ and estimated phase angles $\{\hat{\theta}_i\}_{i = 1}^N$}} 
\caption{Iterative multi-frequency generalized power method}
\label{alg:MF-GPM}
\end{algorithm}

In addition to the spectral method based on the MF-CPQR factorization proposed in Section~\ref{sec:MFCPQR}, we develop an iterative multi-frequency generalized power method for the joint estimation problem, which is inspired by the generalized power method~\cite{chen2021non} and the\textit{``multi-frequency''} nature of the MLE formulation~\eqref{eq:cost4}. 

\subsection{Detailed Steps and Motivations}

Since the joint estimation problem is non-convex, the iterative multi-frequency generalized power method requires a good initialization of the cluster structure and associated phase angles that are sufficiently close to the ground truth. Various spectral algorithms (e.g., CPQR-type algorithm~\cite[Algorithm 1]{fan2021spectral}, ~\cite[Algorithm 3]{chen2021non}, and Algorithm~\ref{alg:spec_alg}) can be used for initialization. 
It is observed experimentally that random initialization will result in convergence to a sub-optimal solution.  Each iteration of Algorithm~\ref{alg:MF-GPM} consists of three main steps. The first step (line 3) is the matrix multiplication between $\bm{A}^{(k)}$ and $\widehat{\bm{V}}^{(k),t}$ for all $k=-K_\text{max},\ldots, K_\text{max}$ (line 4). Then we leverage (line 4) $\widehat{\bm{V}}^{(k),t+1}$ across all frequencies to aggregate and refine the information needed for the joint estimation problem~\eqref{eq:v_hat}, which is inspired by~\eqref{eq:cluster1}. The last step is estimating the cluster structure and associated phase angles. As mentioned before, giving $\widehat{\bm{V}}^{\text{max}, t+1}$ and then finding the corresponding cluster assignment is equal to solving the MCAP (see Definition~\ref{def:proj_H}). This is equivalent to projecting $\widehat{\bm{V}}^{\text{max}, t+1}$ onto the feasible set $\mH$ (line 5), after which the matrix $\widehat{\bm{H}}^{t+1}$ is obtained. The reason why the projection $\mathcal{P}(\cdot)$ is needed rather than directly using the index of the largest entry in each row of $\widehat{\bm{V}}^{\text{max}, t+1}$ is because the solution of the latter approach does not necessarily satisfy the constraint based on the size of each cluster. The associated phase angles can be recovered according to the recovered cluster structure~\eqref{eq:phase2}. Besides, the modification of the iterative MF-GPM for the joint estimation problem with continuous phase angles is the same as that of the spectral method based on the MF-CPQR factorization.

The iterative GPM in~\cite{liu2017estimation} is built upon the classical power method, which is used to compute the leading eigenvectors of a matrix. The method in~\cite{liu2017estimation} adds an important step: projection onto the feasible set that is induced by the constraints on the cluster structure and phase angles. The iterative MF-GPM introduced here takes a step further by not only taking advantage of the efficiency of the power method and the projection, but also leveraging the information across multiple frequencies. In Section~\ref{sec:exp}, numerical experiments show that the iterative MF-GPM largely outperforms GPM~\cite{chen2021non}.

\subsection{Computational Complexity}
\label{subsec:GPM_complexity}
\begin{table}
	\caption{The computational complexity of Algorithm~\ref{alg:MF-GPM} in each step.}
	\centering
	\begin{tabular}{l||c|}
		\hline\hline
		Steps & Computational Complexity\\ \hline
		1. Initialization &  $\mO(|\mE|)$     \\ \cline{2-2}
		2. Matrix multiplication &  $\mO(K_\text{max}|\mE|)$\\ \cline{2-2}
		3. Combine information & $\mO(NK_\text{max}\log{K_\text{max}})$\\
		\cline{2-2}
		4. Estimation & $\mO(N\log{N})$\\
		\hline
		Total complexity &  $\mO(K_\text{max}|\mE|+N(\log{N}+K_\text{max}\log{K_\text{max}}))$\\
		\hline\hline
	\end{tabular}
	\label{table:comp2}
\end{table}

In this section, we compute the complexity of Algorithm~\ref{alg:MF-GPM} step by step in Table~\ref{table:comp2}. Again, here we assume $M=\Theta(1)$. In terms of initialization, the CPQR-type algorithm~\cite{fan2021spectral} is $\mO(|\mE|)$. The matrix multiplication step consists of $\mO(K_\text{max})$ times of matrix multiplication ($\mO(|\mE|)$ per time). In order to combine information across multiple frequencies, we need to compute $MN$ times of FFT of length-$K_\text{max}$ vectors ($\mO(K_\text{max}\log{K_\text{max}})$ per vector). For estimating cluster structure and associated phase angles, we first need to project $\widehat{\bm{V}}^{\text{max}, t+1}$ onto $\mH$, which is $\mO(N\log N)$. Then complexity of estimating the cluster structure and associated phase angles using $\widehat{\bm{H}}^{t+1}$ is negligible. When the network $\mG$ is densely connected with $|\mE|=\mO(N^2)$, Algorithm~\ref{alg:MF-GPM} ends up with $\mO(K_\text{max}N^2)$ if $N>\log{K_\text{max}}$. However, if $|\mE|=o(N^2)$, for example $\mO(N\log{N})$ and $\mO(N)$, the complexity will be reduced to $\mO(K_\text{max}N\max\{\log{N},\log{K_\text{max}}\})$ and $\mO(N\max\{\log{N}, K_\text{max}\log{K_\text{max}}\})$, respectively. As a result, the computational complexity of Algorithm~\ref{alg:MF-GPM} is very similar to Algorithm~\ref{alg:spec_alg}.

\section{Numerical Experiments}
\label{sec:exp}

\begin{figure}[ht!]
    \centering

    \subfloat[\footnotesize{SRER, CPQR}]{\includegraphics[width = 0.48\columnwidth, trim={0.75cm 0.5cm 0.75cm 0.75cm},clip]{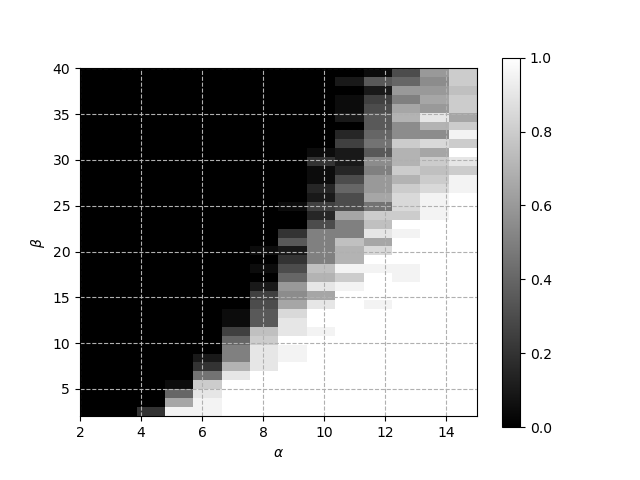}
    \label{fig:SFCPQR_exact}
    }
    \subfloat[\footnotesize{EPS, CPQR}]{\includegraphics[width = 0.48\columnwidth, trim={0.75cm 0.5cm 0.75cm 0.75cm},clip]{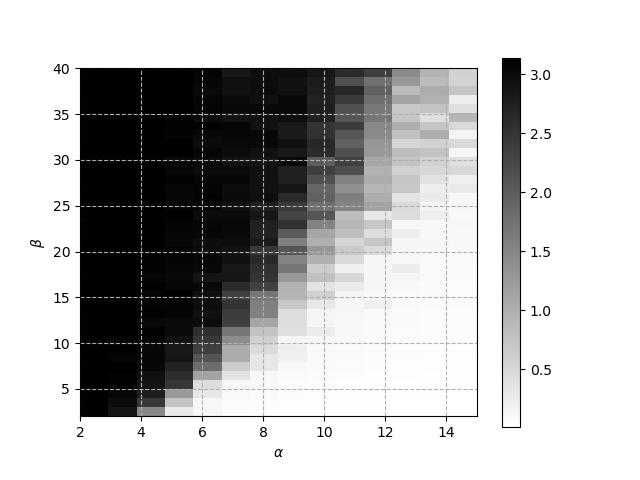}
    \label{fig:SFCPQR_angle}
    }\\
    \subfloat[\footnotesize{SRER, MF-CPQR}]{\includegraphics[width = 0.48\columnwidth, trim={0.75cm 0.5cm 0.75cm 0.75cm},clip]{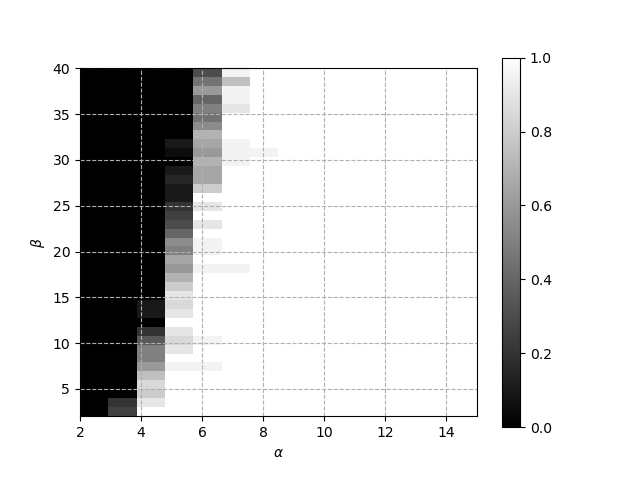}
    \label{fig:MFCPQR_exact}}
    \subfloat[\footnotesize{EPS, MF-CPQR}]{\includegraphics[width = 0.48\columnwidth, trim={0.75cm 0.5cm 0.75cm 0.75cm},clip]{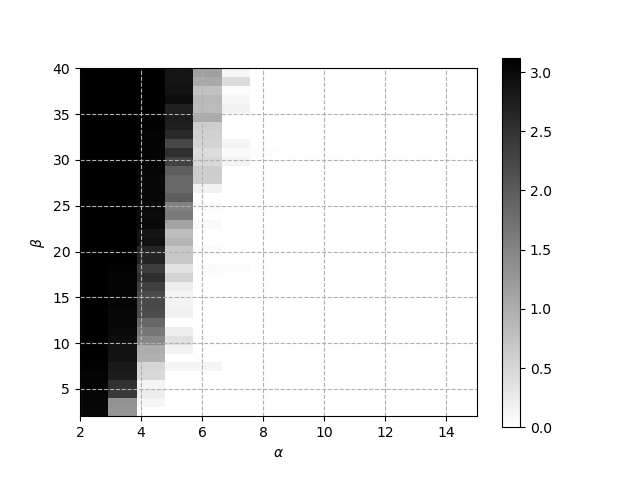}\label{fig:MFCPQR_angle}}
 
    \caption{Comparison 
    between the CPQR-type algorithm~\cite{fan2021spectral} (in the first row) and the spectral method based on the MF-CPQR factorization (in the second  row) in terms of the success rate of exact recovery (SRER) and the error of phase synchronization (EPS), where a smaller black area in each figure indicates a better performance. Experiments are conducted with the setting $M=2$, $N=1000$, and $K_\text{max}=16$. \protect \subref{fig:SFCPQR_exact} and \protect \subref{fig:MFCPQR_exact}: SRER~\eqref{eq:success_rate} under varying $\alpha$ in  $p=\alpha\sfrac{\log{N}}{N}$ and $\beta$ in $q=\beta\sfrac{\log{N}}{N}$; \protect \subref{fig:SFCPQR_angle} and \protect \subref{fig:MFCPQR_angle}: EPS~\eqref{eq:error_rate} under varying $\alpha$ and $\beta$.
    }
    \label{fig:exp1}
\end{figure}

\begin{figure}[ht!]
    \centering

    \subfloat[\scriptsize{SRER, GPM}]{\includegraphics[width = 0.485\columnwidth, trim={0.75cm 0.5cm 0.75cm 0.75cm},clip]{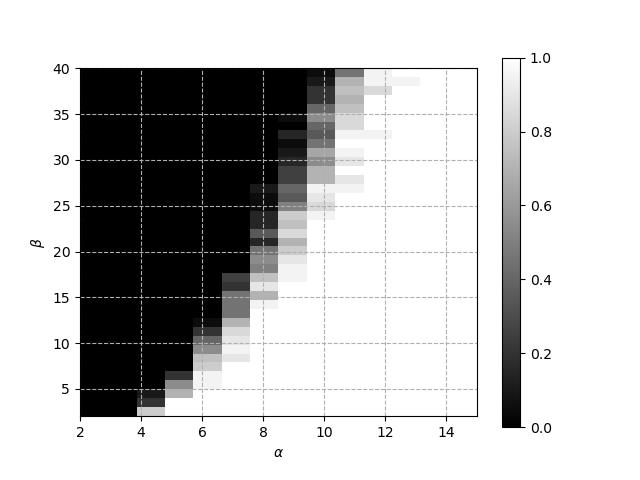}
    \label{fig:SFGPM_exact}
    }
    \subfloat[\scriptsize{EPS, GPM}]{\includegraphics[width = 0.485\columnwidth, trim={0.75cm 0.5cm 0.75cm 0.75cm},clip]{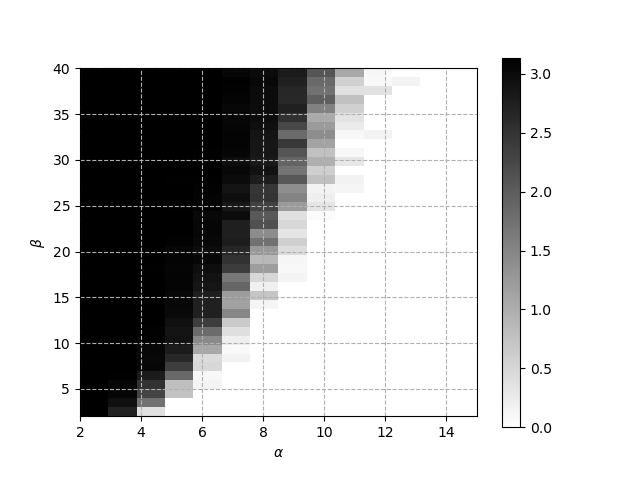}
    \label{fig:SFGPM_angle}
    }\\
    \subfloat[\scriptsize{SRER, MF-GPM}]{\includegraphics[width = 0.485\columnwidth, trim={0.75cm 0.5cm 0.75cm 0.75cm},clip]{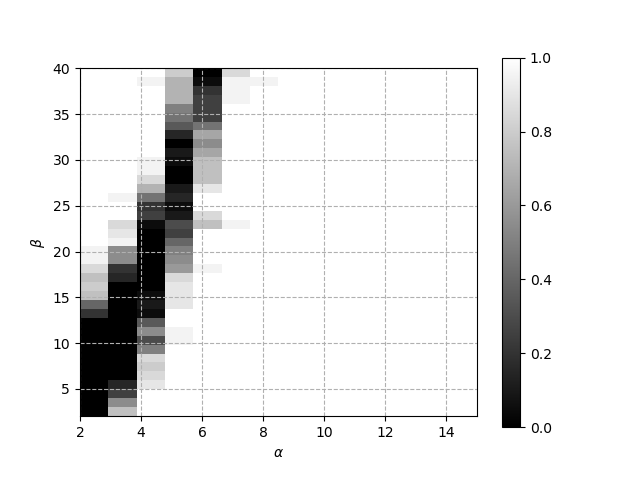}
    \label{fig:MFGPM_exact}}
    \subfloat[\scriptsize{EPS, MF-GPM}]{\includegraphics[width = 0.485\columnwidth, trim={0.75cm 0.5cm 0.75cm 0.75cm},clip]{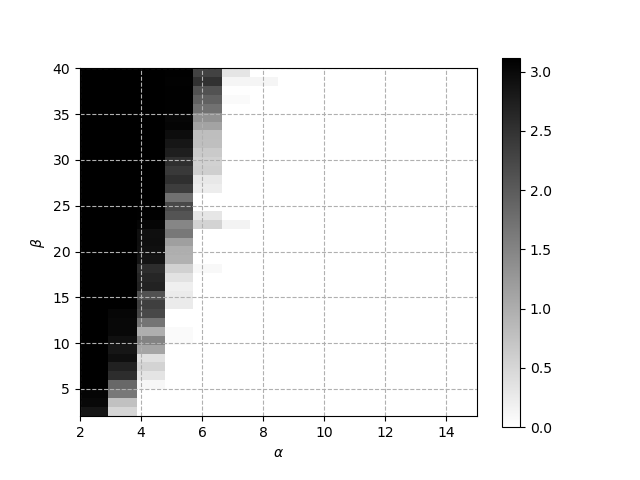}
    \label{fig:MFGPM_angle}}
 
    \caption{Comparison 
    between the GPM~\cite{chen2021non} (in the first row) and the iterative MF-GPM (in the second  row) in terms of the success rate of exact recovery (SRER) and the error of phase synchronization (EPS), where a smaller black area in each figure indicates a better performance. Experiments are conducted with the same setting as Figure~\ref{fig:exp1}. \protect \subref{fig:SFGPM_exact} and \protect \subref{fig:MFGPM_exact}: SRER~\eqref{eq:success_rate} under varying $\alpha$ in  $p=\alpha\sfrac{\log{N}}{N}$ and $\beta$ in $q=\beta\sfrac{\log{N}}{N}$; \protect \subref{fig:SFGPM_angle} and \protect \subref{fig:MFGPM_angle}: EPS~\eqref{eq:error_rate} under varying $\alpha$ and $\beta$.
    }
    \label{fig:exp2}
\end{figure}

\begin{figure*}[ht!]
    \centering

    \subfloat[\scriptsize{SRER, CPQR}]{\includegraphics[width = 0.24\textwidth, trim={0.75cm 0.5cm 0.75cm 0.75cm},clip]{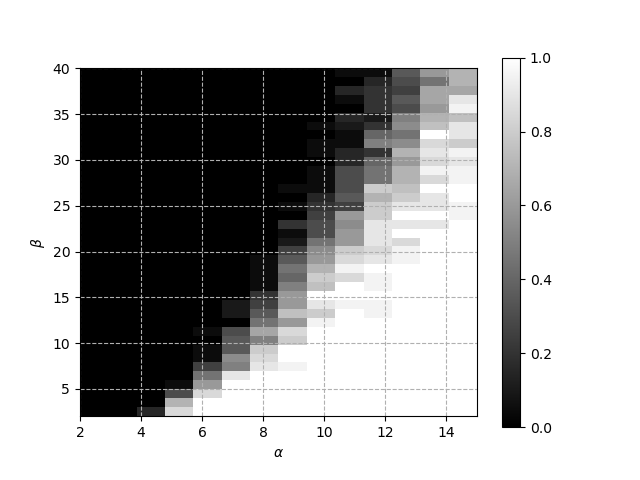}
    \label{fig:SFCPQR_exact_cont}
    }
    \subfloat[\scriptsize{EPS, CPQR}]{\includegraphics[width = 0.24\textwidth, trim={0.75cm 0.5cm 0.75cm 0.75cm},clip]{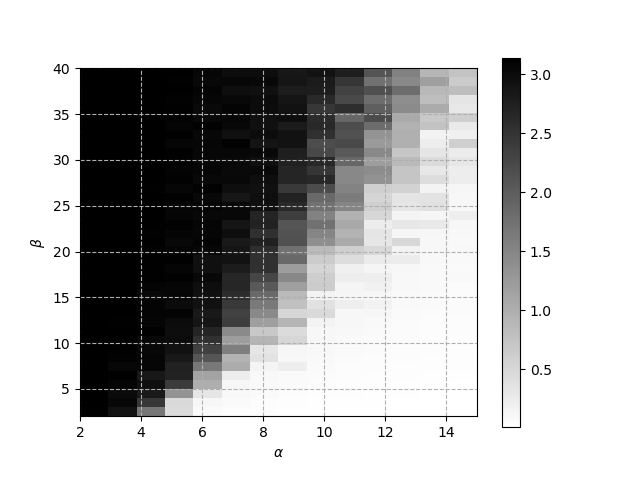}
    \label{fig:SFCPQR_angle_cont}
    }
    \subfloat[\scriptsize{SRER, GPM}]{\includegraphics[width = 0.24\textwidth, trim={0.75cm 0.5cm 0.75cm 0.75cm},clip]{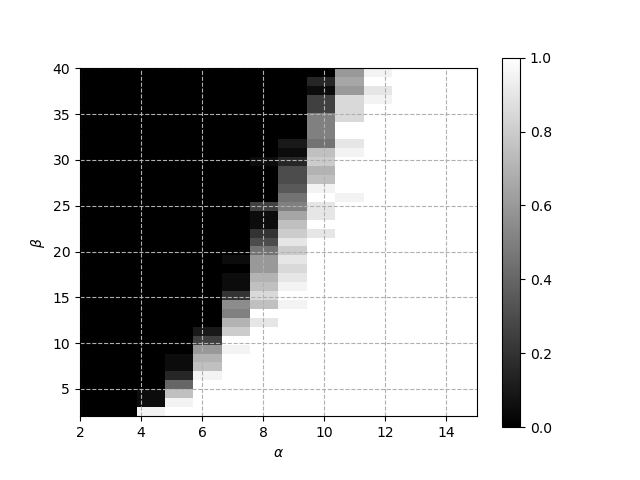}
    \label{fig:SFGPM_exact_cont}}
    \subfloat[\scriptsize{EPS, GPM}]{\includegraphics[width = 0.24\textwidth, trim={0.75cm 0.5cm 0.75cm 0.75cm},clip]{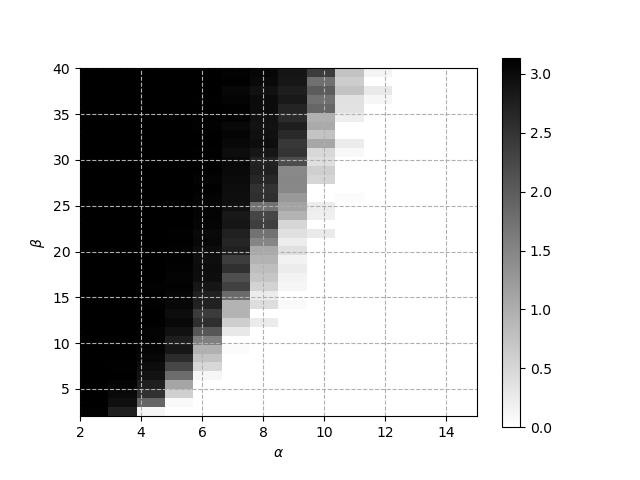}
    \label{fig:SFGPM_angle_cont}} \\

    \subfloat[\scriptsize{SRER, MF-CPQR-5}]{\includegraphics[width = 0.24\textwidth, trim={0.75cm 0.5cm 0.75cm 0.75cm},clip]{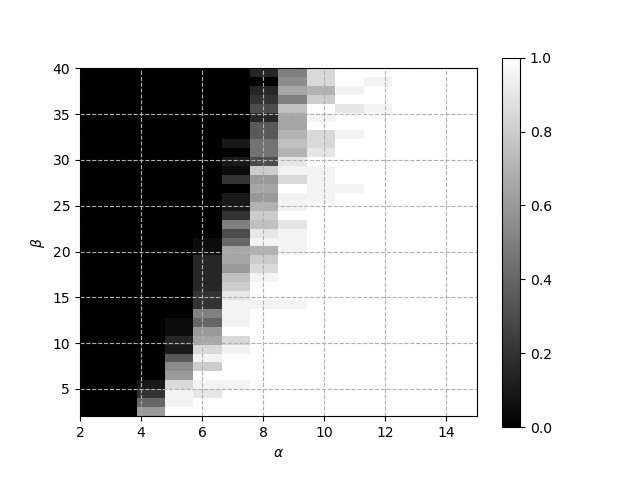}
    \label{fig:MFCPQR_exact_cont_K_5}
    }
    \subfloat[\scriptsize{EPS, MF-CPQR-5}]{\includegraphics[width = 0.24\textwidth, trim={0.75cm 0.5cm 0.75cm 0.75cm},clip]{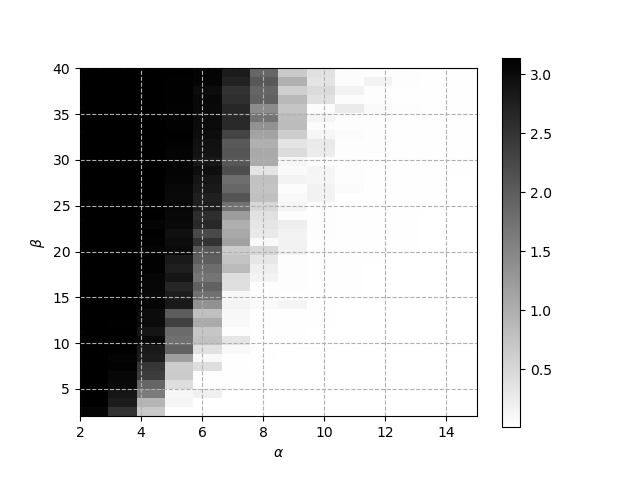}
    \label{fig:MFCPQR_angle_cont_K_5}
    }
    \subfloat[\scriptsize{SRER, MF-GPM-5}]{\includegraphics[width = 0.24\textwidth, trim={0.75cm 0.5cm 0.75cm 0.75cm},clip]{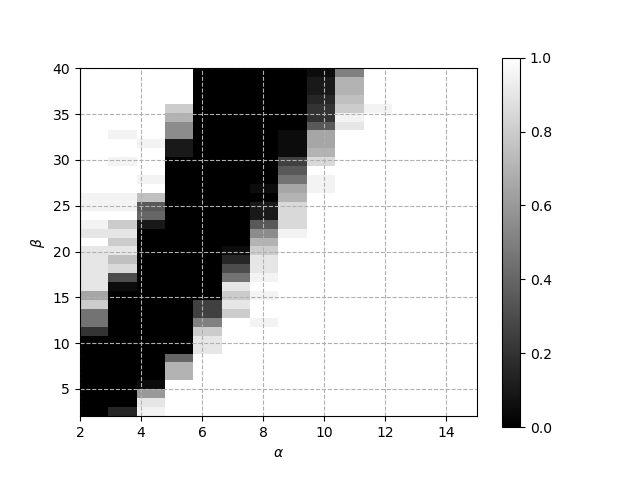}
    \label{fig:MFGPM_exact_cont_K_5}}
    \subfloat[\scriptsize{EPS, MF-GPM-5}]{\includegraphics[width = 0.24\textwidth, trim={0.75cm 0.5cm 0.75cm 0.75cm},clip]{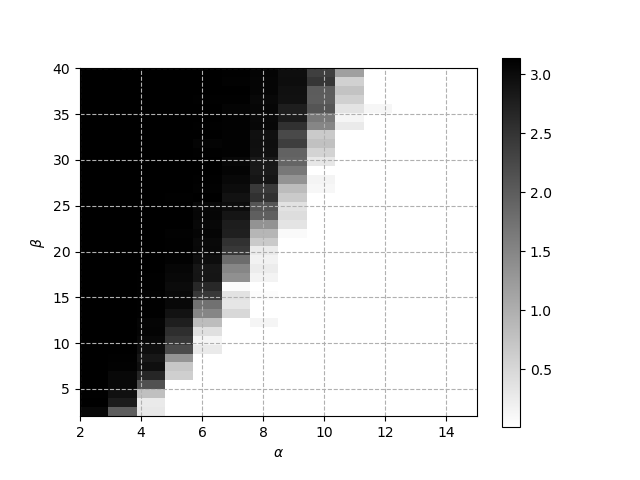}
    \label{fig:MFGPM_angle_cont_K_5}} \\
    
    \subfloat[\scriptsize{SRER, MF-CPQR-10}]{\includegraphics[width = 0.24\textwidth,trim={0.75cm 0.5cm 0.75cm 0.75cm},clip]{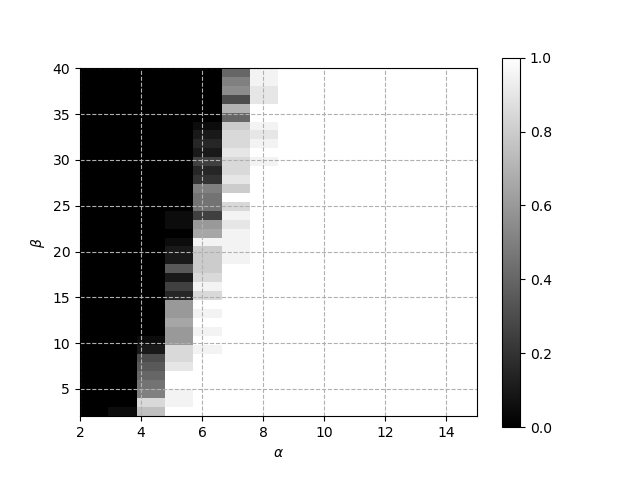}
    \label{fig:MFCPQR_exact_cont_K_10}
    }
    \subfloat[\scriptsize{EPS, MF-CPQR-10}]{\includegraphics[width = 0.24\textwidth, trim={0.75cm 0.5cm 0.75cm 0.75cm},clip]{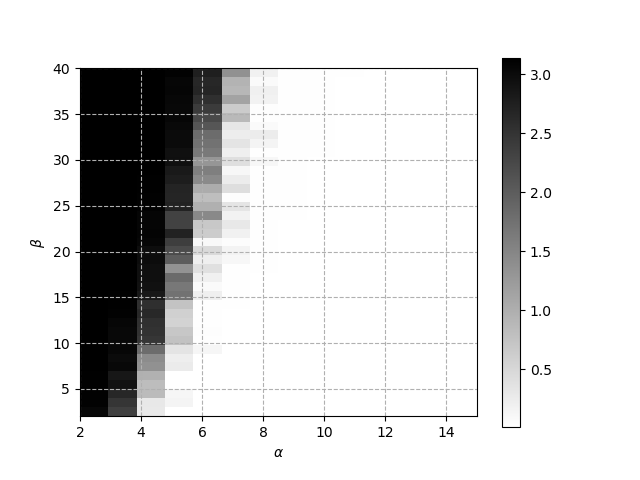}
    \label{fig:MFCPQR_angle_cont_K_10}
    }
    \subfloat[\scriptsize{SRER, MF-GPM-10}]{\includegraphics[width = 0.24\textwidth, trim={0.75cm 0.5cm 0.75cm 0.75cm},clip]{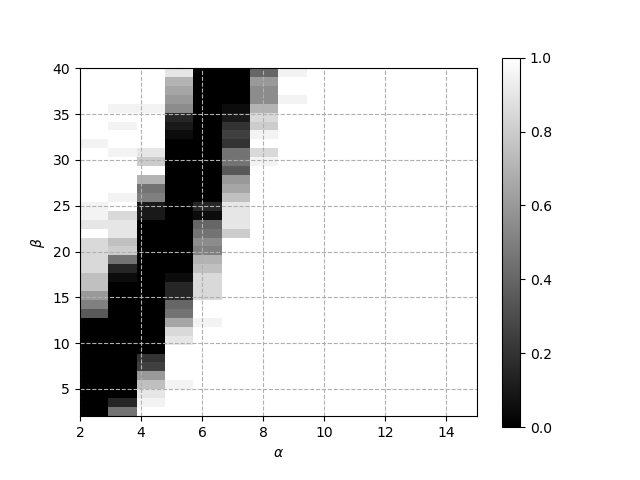}
    \label{fig:MFGPM_exact_cont_K_10}}
    \subfloat[\scriptsize{EPS, MF-GPM-10}]{\includegraphics[width = 0.24\textwidth, trim={0.75cm 0.5cm 0.75cm 0.75cm},clip]{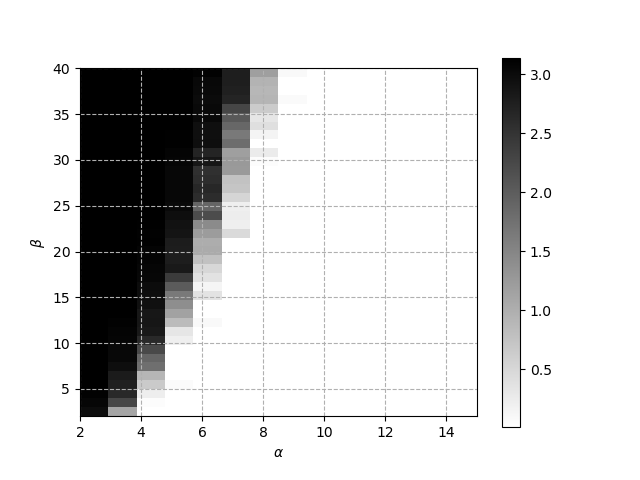}
    \label{fig:MFGPM_angle_cont_K_10}}\\
    
    \subfloat[\scriptsize{SRER, MF-CPQR-20}]{\includegraphics[width = 0.24\textwidth, trim={0.75cm 0.5cm 0.75cm 0.75cm},clip]{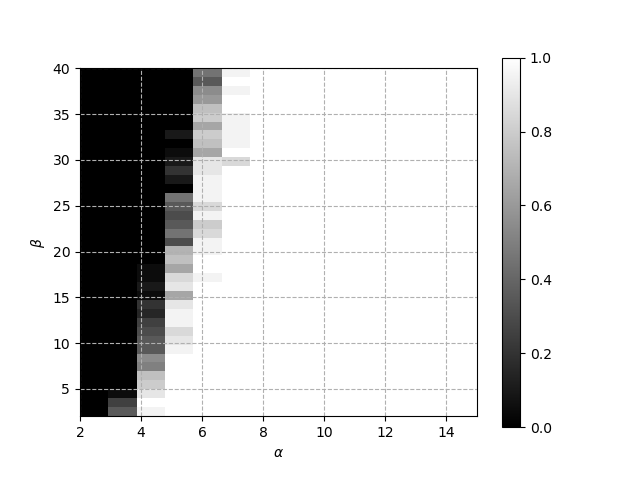}
    \label{fig:MFCPQR_exact_cont_K_20}}
    \subfloat[\scriptsize{EPS, MF-CPQR-20}]{\includegraphics[width = 0.24\textwidth, trim={0.75cm 0.5cm 0.75cm 0.75cm},clip]{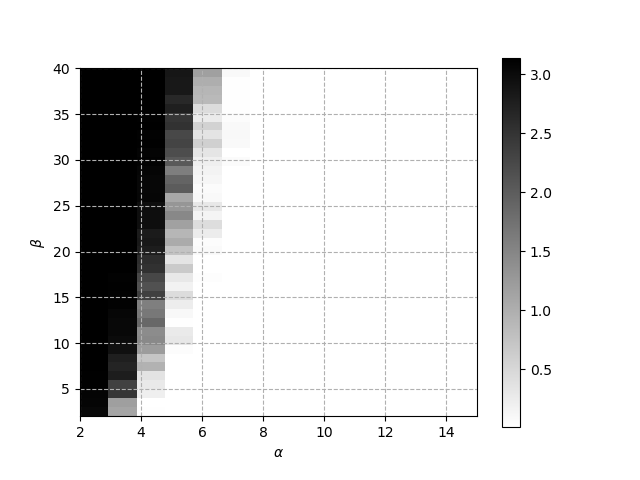}
    \label{fig:MFCPQR_angle_cont_K_20}}
    \subfloat[\scriptsize{SRER, MF-GPM-20}]{\includegraphics[width = 0.24\textwidth, trim={0.75cm 0.5cm 0.75cm 0.75cm},clip]{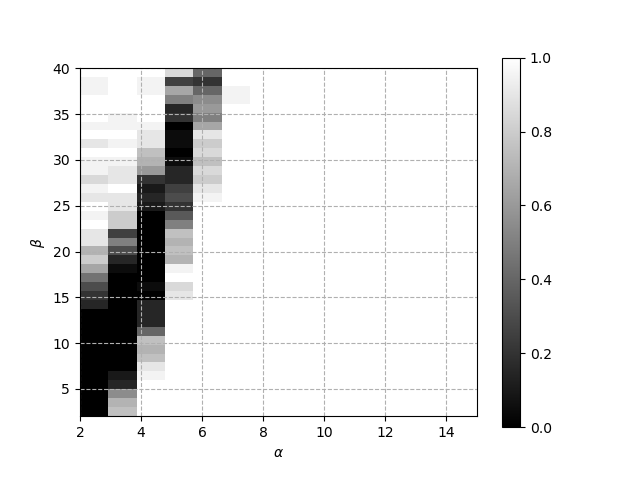}
    \label{fig:MFGPM_exact_cont_K_20}}
    \subfloat[\scriptsize{EPS, MF-GPM-20}]{\includegraphics[width = 0.24\textwidth, trim={0.75cm 0.5cm 0.75cm 0.75cm},clip]{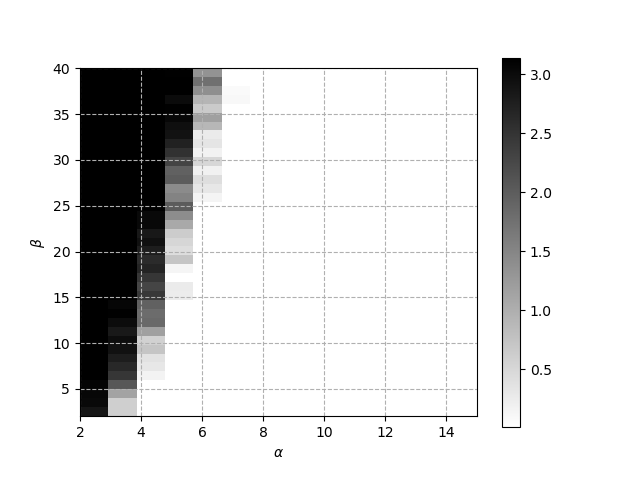}
    \label{fig:MFGPM_angle_cont_K_20}}
 
    \caption{Results for the joint estimation problem with continuous phase angles in $[0,2\pi)$ using the CPQR-type algorithm~\cite{fan2021spectral}, the GPM~\cite{chen2021non}, the spectral method based on the MF-CPQR factorization, and the iterative MF-GPM, where a smaller black area in each figure indicates better performance. The choice of $M$ and $N$ are the same as Figure~\ref{fig:exp1}. The first and third columns show the SRER, and the second and fourth columns shows EPS. \protect  \subref{fig:SFCPQR_exact_cont}, \protect  \subref{fig:SFCPQR_angle_cont}, \protect  \subref{fig:SFGPM_exact_cont}, and \protect  \subref{fig:SFGPM_angle_cont}: The
    results of the CPQR-type algorithm~\cite{fan2021spectral} and the GPM~\cite{chen2021non}; \protect  \subref{fig:MFCPQR_exact_cont_K_5}, \protect  \subref{fig:MFCPQR_angle_cont_K_5}, \protect  \subref{fig:MFGPM_exact_cont_K_5}, and \protect  \subref{fig:MFGPM_angle_cont_K_5}: The results of the spectral method based on the MF-CPQR factorization and the iterative MF-GPM with $K_\text{max}=5$; \protect  \subref{fig:MFCPQR_exact_cont_K_10}, \protect  \subref{fig:MFCPQR_angle_cont_K_10}, \protect  \subref{fig:MFGPM_exact_cont_K_10}, and \protect  \subref{fig:MFGPM_angle_cont_K_10}: The results of the spectral method based on the MF-CPQR factorization and the iterative MF-GPM with $K_\text{max}=10$; \protect  \subref{fig:MFCPQR_exact_cont_K_20}, \protect  \subref{fig:MFCPQR_angle_cont_K_20}, \protect  \subref{fig:MFGPM_exact_cont_K_20}, and \protect  \subref{fig:MFGPM_angle_cont_K_20}: The results of the spectral method based on the MF-CPQR factorization and the iterative MF-GPM with $K_\text{max}=20$.}
    \label{fig:exp3}
\end{figure*}

This section deals with numerical experiments of the spectral method based on the MF-CPQR factorization (Algorithm~\ref{alg:spec_alg}) and the iterative MF-GPM (Algorithm~\ref{alg:MF-GPM}) to showcase their performance against state-of-the-art benchmark algorithms\footnote{Codes are available at \url{https://github.com/LingdaWang/Joint_Community_Detection_and_Phase_Synchronization}}. For comparison, the benchmark algorithms are chosen as i) the CPQR-type algorithm~\cite{fan2021spectral}, ii) the GPM~\cite{chen2021non}, where both of them can be modified identically from the joint community and group synchronization problem into the joint community detection and phase synchronization problem. Specifically, algorithms in~\cite{fan2021spectral,chen2021non} are single frequency version of our proposed algorithms, which can be realized by replacing the summation over $k$ in~\eqref{eq:cluster1},~\eqref{eq:phase1},~\eqref{eq:v_hat}, and~\eqref{eq:phase2} with $k=1$. 

In each experiment, we generate the observation matrix $\bm{A}$ using the probabilistic model, SBM-Ph, as discussed in Section~\ref{sec:Problem_Formulation} and estimate the cluster structure and associated phase angles by the spectral algorithms based on the MF-CPQR factorization, the iterative MF-GPM, and the benchmark algorithms. To evaluate the numerical results, we defined two metrics, \textit{success rate of exact recovery} (SRER) and \textit{error of phase synchronization} (EPS), for recovering the cluster structure and associated phase angles. In terms of SRER, it shows the rate of algorithms exactly recover the cluster structure. Let $\hat{\mS}_{m}=\{i\in[N]|\hat{\mM}(i)=m\}$ be the set of nodes assigned into the $m$th cluster by algorithms, and we have that
\begin{equation}
\label{eq:success_rate}
\begin{aligned}
     \textup{SRER} = \textup{ the rate } \{\hat{\mS}_{m}\}_{m=1}^M \textup{ is identical to } \{\mS_{m}\}_{m=1}^M.   
\end{aligned}
\end{equation}
As for the EPS, it assesses the performance of recovering phase angles. We define $\bm{\theta}^{*,(m)}=[e^{\iota\theta_i^*}]_{i\in\mS^*_m}\in\C^{s}$ for each cluster that concatenates the ground truth $\theta^*_i$ for all $i\in\mS^*_m$, and similarly $\hat{\bm{\theta}}^{(m)}=[e^{\iota\hat{\theta}_i}]_{i\in\mS^*_m}\in\C^{s}$ for the estimated phase angles. Then, after removing the ambiguity with aligning $\hat{\bm{\theta}}^{(m)}$ with $\bm{\theta}^{*,(m)}$ in each cluster as
\begin{equation*}
\begin{aligned}
&\gamma^{(m)} = \argmin_{g^{(m)}\in\Omega \text{ or } [0,2\pi)}\|\hat{\bm{\theta}}^{(m)}e^{\iota g^{(m)}}-\bm{\theta}^{*,(m)}\|_2, \\
&\quad\forall m=1,\ldots, M,   
\end{aligned}
\end{equation*}
the EPS is defined as
\begin{equation}
\label{eq:error_rate}
\begin{aligned}
&\textup{EPS}= \\
&\max_{m\in[M]}\max_{i\in\mS^*_m}\{\min (|\hat{\theta}_i+\gamma^{(m)}-\theta^*_i|, 2\pi - |\hat{\theta}_i+\gamma^{(m)}-\theta^*_i|)\}.
\end{aligned}
\end{equation}
The EPS is actually the maximum error of estimated phase angles among all nodes. Besides, both the SRER and EPS are computed over 20 independent and identical realizations for each experiment in the following. In the rest of this section, we first present the results of the joint estimation problem in Section~\ref{subsec:exp1-1} and followed by the extension to continuous phase angles in Section~\ref{subsec:exp1-2}.

\subsection{Results of the Joint Estimation Problem}
\label{subsec:exp1-1}

We first show the results of the spectral method based on the MF-CPQR factorization (Algorithm~\ref{alg:spec_alg}) against the CPQR-type algorithm~\cite{fan2021spectral} on the joint estimation problem, where the case of $M=2$, $s=500$, and $K_\text{max}=16$ is considered. Similar to~\cite{fan2021spectral,chen2021non}, we test the recovery performance in the regime $p,q=\mO(\sfrac{\log{N}}{N})$, where different $p=\alpha\sfrac{\log{N}}{N}$ and $q=\beta\sfrac{\log{N}}{N}$ with varying $\alpha$ and $\beta$ are included. In Figure~\ref{fig:exp1}, we show SRER~\eqref{eq:success_rate} and EPS~\eqref{eq:error_rate}. As one can observe from Figure~\ref{fig:SFCPQR_exact} and~\ref{fig:MFCPQR_exact}, our proposed spectral method based on the MF-CPQR factorization outperforms the CPQR-type algorithm~\cite{fan2021spectral} in SRER. EPS follows a similar pattern. 

Next, we test the performance of the iterative MF-GPM (Algorithm~\ref{alg:MF-GPM}) against the GPM~\cite{chen2021non} under the same choice of $M$, $s$, and $K_\text{max}$ as before. Since the GPM and the iterative MF-GPM require initialization that is close enough to the ground truth, we can choose either~\cite[Algorithm 3]{chen2021non} or the CPQR-type algorithm~\cite{fan2021spectral}. We set the number of iterations to be $50$ as suggested by~\cite{chen2021non}. Again, as one can observe from Figure~\ref{fig:exp2}, our proposed iterative MF-GPM achieves higher accuracy in both SRER and EPS. Surprisingly, one may also notice the region where $p$ is small and $q$ is large (top left area in Figure~\ref{fig:MFGPM_exact}), the iterative MF-GPM is capable of recovering the cluster structure with high probability, however, this is not the case in recovering associated phase angles. 

When compare the results shown in Figure~\ref{fig:exp1} and~\ref{fig:exp2} together, the spectral method based on the MF-CPQR factorization shows very similar result as the iterative MF-GPM, which are both significantly better than the GPM~\cite{chen2021non} and the CPQR-type algorithm~\cite{fan2021spectral}. However, compared to the iterative MF-GPM, the spectral method based on the MF-CPQR factorization is free of initialization. One may also observe the performance of the GPM~\cite{chen2021non} outperform the CPQR-type algorithm~\cite{fan2021spectral}.

\subsection{Results with Continuous Phase Angles}
\label{subsec:exp1-2}

In this section, we show the results of our proposed algorithms against benchmark algorithms on the joint estimation problem with continuous phase angles. As mentioned in Section~\ref{subsec:ext}, the algorithms tested in Section~\ref{subsec:exp1-1} can be directly applied after simple modification (See Section~\ref{subsec:motivation} for details), and thus we choose the similar setting as Section~\ref{subsec:exp1-1}. Besides, since~\eqref{eq:cont_cost3} is a truncated MLE formulation of the true one~\eqref{eq:MLE2}, experiments of the spectral method based on the MF-CPQR factorization and the iterative MF-GPM with different $K_\text{max}$ are conducted to study the trend of the results as $K_\text{max}$ grows. The results are detailed in Figure~\ref{fig:exp3}, with very similar performance as shown in Figure~\ref{fig:exp1} and~\ref{fig:exp2}. In addition, as $K_\text{max}$ grows, the cluster structure recovery and phase synchronization become more accurate in both MF-CPQR based method and iterative MF-GPM. 

To choose a suitable $K_\text{max}$ for the continuous phase angles, we need to consider the trade-off between the performance and the computational complexity. We observe that the estimation accuracy is improved as $K_\text{max}$ increases. On the other hand, the computational complexity scales linearly with $K_\text{max}$. 
In addition,  the computational complexity also depends on the number of nodes $N$ and the number of clusters $M$, which needs to be taken into consideration for the trade-off between accuracy and efficiency. 
Thus, it is difficult to state a simple optimal policy for choosing $K_\text{max}$ for the continuous phase angles. Despite this, we have shown that our methods outperform the CPQR-type algorithm and the GPM as long as $K_\text{max}\ge 1$, and moreover largely outperform other baseline algorithms when $K_\text{max}\ge 10$. Therefore, our choice of $K_\text{max}$ is between $10$ to $30$ for most cases.

\section{Conclusion}
\label{sec:conclusion}

In this paper, we study the joint community detection and phase synchronization problem from an MLE perspective, and provide the new insight that its MLE formulation has a \textit{``multi-frequency''} nature. We then propose two methods, the spectral method based on the novel MF-CPQR factorization and the iterative MF-GPM, to tackle the MLE formulation of the joint estimation problem, where the latter one requires the initialization from spectral methods. Numerical experiments demonstrate the advantage of our proposed algorithms against state-of-the-art algorithms.

It remains open to establish the theoretical analysis that can tightly characterize the noise robustness of our proposed algorithms. Sub-optimal bounds can be easily derived following the analysis in~\cite{fan2021spectral,chen2021non} by considering the frequency-1 component. However, these results do not explore additional frequency information. The key difficulties lie in i) analyzing the properties and relationships of eigenvectors among different frequency components with dependent noises, and ii) analyzing the power method across multiple frequencies. We leave them for future investigation.  

In addition, there are several directions that can be further explored. It is natural to expect that the proposed approach can be extended to compact non-Abelian groups (e.g., rotational groups, orthogonal groups, and symmetric groups) using the corresponding irreducible representations.

\section*{Acknowledgments}
We would like to thank Dr. Xiuyuan Cheng, and Yifeng Fan for helpful discussions.

\bibliographystyle{IEEEtran}
\bibliography{refs.bib}

\end{document}